\documentclass[
journal=jctcce,
manuscript=article,
layout= twocolumn% traditional or twocolumn
]{achemso}

\usepackage{mathtools,amsmath,amsfonts,stmaryrd,amssymb}
\usepackage{siunitx}
\usepackage[version=3]{mhchem} % Formula subscripts using \ce{} Typeset chemical formulae/equations and H and P statements
\usepackage{caption}
\usepackage{color, colortbl}

\usepackage{caption, graphicx, grffile, wrapfig, subcaption}
\usepackage{array, booktabs} %for creating tables
\usepackage{float}
\graphicspath{ {./figs/} }
\DeclareGraphicsExtensions{.pdf,.png,.jpg,.eps}

\newcommand*{\mean}[1]{\left< #1 \right>}

\newcommand*{\down}{\textsubscript}

\newcommand{\kb}{\mathop{}\! k_\mathrm{B}}
\newcommand*{\kT}{\mathop{}\! k_\mathrm{B} T}
\newcommand*{\diff}{\mathop{}\!\mathrm{d}}

\DeclareSymbolFont{rmlargesymbols}{OMX}{mdbch}{m}{n}
\DeclareMathSymbol{\rmintop}{\mathop}{rmlargesymbols}{82}

\newcommand{\SW}[1]{{\color{black}#1}}
\newcommand{\SIStbsystem}{S1}
\newcommand{\SIStbsim}{S2}
\newcommand{\SIStbsimRMSD}{S3}
\newcommand{\SIStbsimPCA}{S4}
\newcommand{\SIStbRMSDsankey}{S5}
\newcommand{\SIStbPCAsankey}{S6}
\newcommand{\SIStbRMSDsilhouette}{S7}
\newcommand{\SIStbPCAsilhouette}{S8}
\newcommand{\SIAdeSystem}{S9}
\newcommand{\SIAdesim}{S10}
\newcommand{\SIAdefel}{S11}
\newcommand{\SIAdeRMSDsimblock}{S11}
\newcommand{\SIAdeLeidensankey}{S13}
\newcommand{\SIAdeLeidenfelbootstrapp}{S12}
\newcommand{\SIAdekmedsilhouette}{S14}

\newcommand{\SIAdeAllsankey}{S16}

\newcommand{\SIAdeNeighbor}{S20}

\newcommand{\SIAdeinvlipid}{S22}
\newcommand{\SIAdecontacts}{S23}

\usepackage{todonotes}
\makeatletter
\renewcommand{\todo}[2][]{%
    \@todo[caption={#2}, #1]{\begin{spacing}{0.5}#2\end{spacing}}%
} 
\makeatother

\author{Victor Tänzel}
\affiliation{Biomolecular Dynamics, Institute of Physics, University of Freiburg, 79104 Freiburg}
\altaffiliation{authors contributed equally}
\alsoaffiliation{present address: Statistical Physics of Soft Matter and Complex Systems, Institute of Physics, University of Freiburg, 79104 Freiburg}
\author{Miriam Jäger}
\affiliation{Biomolecular Dynamics, Institute of Physics, University of Freiburg, 79104 Freiburg}
\altaffiliation{authors contributed equally}
\author{Steffen Wolf}
\email{steffen.wolf@physik.uni-freiburg.de}
\affiliation{Biomolecular Dynamics, Institute of Physics, University of Freiburg, 79104 Freiburg}

\title{Learning protein-ligand unbinding pathways via single-parameter community detection}

\abbreviations{dcTMD,MD,TMD,CPM}
\keywords{biased MD simulations, pathways, machine learning, drug unbinding}

\begin{document}

%\begin{tocentry}
%
%    \centering
%    \includegraphics[width=0.8\linewidth]{./figs/TOC_5_lachs.png}
%
%
%\end{tocentry}
%\listoftodos

\begin{abstract}
  Understanding the dynamics of biomolecular complexes, e.g., of protein-ligand (un)binding, requires the understanding of paths such systems take between metastable states. In MD simulations, paths are usually not observable per se, but need to be inferred from simulation trajectories. Here, we present a novel approach to cluster trajectories based on a community detection algorithm that necessitates only the definition of a single parameter. The unbinding of the streptavidin-biotin complex is used as a benchmark system and the A\down{2a} adenosine receptor in complex with the inhibitor ZM241385 as an elaborate application. We demonstrate how such clusters of trajectories correspond to pathways, and how the approach helps in the identification of reaction coordinates for the considered (un)binding process. 
\end{abstract}


\section{Introduction}

The formation and dissociation of protein-ligand complexes are key processes in signal transduction within and between cells. As optimized binding kinetics in the form of long residence times can improve a drug's efficacy, understanding the binding and unbinding of drug molecules to and from the target proteins has become a focus of biophysical and medical research \cite{Copeland2006,Swinney2008,Copeland2016a,Schuetz2017}.

The observation of spontaneous ligand (un)binding in molecular dynamics (MD) simulations poses a challenge for current computing hardware, since the time scales of these processes can reach up to several hours of real time. Accordingly, biased simulation approaches such as infrequent metadynamics\cite{Tiwary2013,Ray2023}, SEEKR\cite{Votapka2017,Ojha2023}, $\tau$RAMD\cite{Kokh2018}, ligand Gaussian accelerated MD\cite{Miao2020}, steered MD\cite{Rico2019,Potterton2019,Cai2023,Iida2023} and others have been developed to efficiently sample (un)binding processes.

Understanding (un)binding events is intrinsically connected to the pathways that a drug takes towards or away from its binding site\cite{Wolf2023a}. To characterize these processes, it is necessary to define such pathways using appropriate collective variables (CVs),
%\vec{y}$
which represent linear or nonlinear combinations of the atomic Cartesian coordinates. Finding the relevant process paths is challenging as they are unknown a priori and can only be inferred from simulation data a posteriori. Similarly, a good reaction coordinate (RC), i.e., a CV that solely encodes the (un)binding process, has to be characterized in post-processing\cite{Tiwary2015,Huang2017,Schuetz2019,Capelli2019,Rydzewski2019,Kokh2020,Bianciotto2021,Bray2022,Motta2022}. Therefore, RCs may differ from the bias coordinate employed to drive (un)binding events. 

For the prediction of both binding and unbinding rates, our group has developed dissipation-corrected targeted Molecular Dynamics (dcTMD)\cite{Wolf2018,Post2023}, which allows estimating free energies $\Delta G$ and friction factors $\Gamma$ from fully atomistic MD simulations with constant velocity constraint. These fields are employed in a numerical integration of a Langevin equation, which are a popular approach to coarse-grain dynamics\cite{Zwanzig2001,Schilling2022}, to effectively sample (un)binding processes and infer their time scales\cite{Wolf2020}. The crucial point for the dissipation correction is the identification of all relevant pathways taken by the biased trajectories\cite{Wolf2023}. To this end, we follow a conceptual framework that is inspired by transition pathway sampling,\cite{Bolhuis2002,Berryman2010} time series analysis\cite{Yuan2017,Ray2023a} and the clustering of protein conformational states\cite{Sittel2018,Nagel2023} by formulating the search for ligand unbinding pathways as a clustering task: we define a set of relevant input features, i.e., a pre-selection of CVs with reduced dimensionality compared to the full phase space, as well as a metric that allows to calculate distances between trajectories in the feature space. We then identify clusters of similar trajectories, i.e., groups with small intra-trajectory distances. Lastly, we postulate that all trajectories contained in a cluster follow the same pathway. 
\SW{As our application of a constant velocity constraint renders the pathway probability distribution undefined, we can neither use transition path sampling\cite{Bolhuis2002,Berryman2010} nor density-based clustering\cite{Kriegel2011} for finding paths. We therefore use the assumption that geometrically similar trajectories follow the same pathway, and that a geometric clustering\cite{Saxena2017} based on distance measures between trajectories can reveal groups of pathways.}

A common problem when searching for paths via trajectory clustering is the need for significant user input\cite{Tiwary2015,Schuetz2019,Capelli2019,Rydzewski2019,Kokh2020,Bianciotto2021,Bray2022,Motta2022}. 
\SW{To alleviate potential user input bias, machine learning approaches have been developed that learn RCs on the fly during biasing\cite{Ribeiro20220,Bertazzo2021,Badaoui2022,Francelanord2023,Chen2023,Jung2023,froehlking2024} (see Ref.~\citenum{Mehdi2024} for a recent review). In our case, we wanted to use a machine learning algorithm that a) operates unsupervised, as no ground truth is available for training, b) is fast to screen as many alternative path cluster scenarios as possible, and c) allows for easy backtracking to refine path RCs from the initial guess of input features. In this article, we employ an unsupervised machine learning approach to cluster trajectories that fulfills these requirements and almost completely eliminates user input:}
 Based on a Leiden community detection algorithm\cite{Traag2011,Diez2022}, it only requires the definition of a single resolution parameter $\gamma$, which can be easily varied to test the robustness of the path identification. This community detection has recently been shown to perform well in the identification of allosteric communication pathways in proteins\cite{Diez2022,Post2021}. We compare trajectories based on ligand root-mean-square displacements (RMSDs) and contact principal component\cite{Ernst2015} distances. 

As a first benchmark, we apply our approach to simulations of the streptavidin-biotin complex\cite{Cai2023} (St-b), where we recover predefined unbinding paths. Next, we demonstrate the approach's capabilities with the challenging example of the A\down{2a} adenosine receptor, which is a member of the pharmacologically relevant membrane protein class of G protein-coupled receptors,\cite{Lebon2011,Dore2011,Segala2016} in complex with the inhibitor ZM241385\cite{Segala2016}. Lastly, we present how our approach can lead towards improved RCs that shed light on the microscopic effects defining pathways.

\section{Theory}
\subsection{Dissipation-corrected \\targeted molecular dynamics}
To set the stage for our interest in identifying (un)binding pathways, we shortly recapitulate the formal basis of dcTMD\cite{Wolf2020}. Targeted MD (TMD)\cite{Schlitter1994} forces a system, e.g., a ligand, along a (here one-dimensional) biasing coordinate $x$ via a constant velocity distance constraint $\Phi(t) = x (t) - (x_0 +v t) \overset{!}{=} 0$ with velocity $v$. The constraint is realized via a constraint force $f = \lambda \frac{\diff{\Phi}}{\diff{x}}$ with a Lagrange multiplier $\lambda$.
From the resulting non-equilibrium work $W(x)=\int_{x_0}^{x}\diff x' f(x')$, the free energy 
$\Delta G(x)$ can be estimated using a cumulant expansion of the Jarzynski equality\cite{Jarzynski2004, Park2004} as
\begin{align} \label{eq:Wdiss_prop_Wvariance}
    \Delta G(x) \approx \mean{W(x)}_N  -
        \frac{1}{2\kT} \mean{\delta W(x)^2}_N.
\end{align}
The brackets $\mean{\cdot}_N$ denote an ensemble average over $N$ statistically independent realizations starting out from a common equilibrium Boltzmann distribution, and $\beta^{-1}=\kT$ with the Boltzmann constant $\kb$ and temperature $T$. Truncating the expansion in Eq.~\eqref{eq:Wdiss_prop_Wvariance} after the second cumulant is exact if the work follows a Gaussian distribution, and we identify the dissipative work
\begin{align} \label{eq:Wdiss}
    W_{\rm diss} = \frac{1}{2\kT} \mean{\delta W(x)^2}_N .
\end{align}
Modeling the process using a Markovian Langevin equation with an external force\cite{Wolf2018}, $\Delta G(x)$ and the friction factor $\Gamma (x)$ can be calculated from $W(x)$ via
\begin{align}
    \label{eq:dcTMD}
    \Delta G (x) &= \left<W(x)\right>_N - W_{\rm diss} (x), \\
    \Gamma (x) &= \frac{1}{v} \, \frac{\mathrm{d}}{\mathrm{d}x} W_{\rm diss} (x) .\label{eq:friction}
\end{align}

\subsection{Pathways}

In earlier works, we have shown that the assumption of a normally distributed $W(x)$ breaks down if a system follows paths in a CV space with path-dependent friction\cite{Wolf2020}. When dcTMD is applied to protein-ligand systems, such pathways in CV space may appear and involve different routes of the ligand through the protein \cite{Wolf2020, Wolf2023}, changing protein conformations\cite{Jaeger2021} or even ligand-internal hydrogen bonds\cite{Bray2022}. In practice, a clear indication of pathways and path-dependent friction is an overestimation of the dissipative work $W_{\rm diss}$\cite{Wolf2023}.

To address this challenge, we recently developed two strategies. In principle, given sufficient sampling, a high dimensional model in the CV space can be constructed\cite{Post2023}. However, given the sampling limitations imposed by the computational cost of fully atomistic MD simulations that are necessary for this approach, we find it more practical to work with a one-dimensional model for protein-ligand complexes instead.

We group pulling trajectories according to their similarity $s$ in the CV space and presume that highly similar trajectories follow the same pathway $k$. These groups must be distinguished by low similarity between and high similarity within them. Furthermore, the trajectory ensemble within a cluster exhibits a Gaussian work distribution\cite{Wolf2023}. Their corresponding subspaces in the CV space and friction factors $\Gamma_k(x)$ characterize them, and we define path-wise free energies\cite{Wolf2023}
\begin{align}
	\Delta G_k (x) = \mean{W(x)}_{N_k} - \frac{1}{2\kT}\mean{\delta W(x)}_{N_k}
	\label{eg:dcTMD_dGpathwise}
\end{align}
for a group containing $N_k$ trajectories. These path-wise $\Delta G_k (x)$ and $\Gamma_k (x)$ can then be employed to determine path-wise (un)binding kinetics\cite{Wolf2020,Wolf2023}.

\subsection{Trajectory clustering}
\subsubsection{Input features and distances}

With the intent of identifying pathways in dcTMD trajectories, we start by comparing unbinding trajectories based on two different sets of input features, i.e., CVs. Both consist of internal coordinates that remove contributions from overall system translation and rotation\cite{Ernst2015,Sittel2018}. The first set of features are protein-aligned Cartesian coordinates, in which we evaluate pairs of ligand trajectories by the root-mean-square displacement (RMSD) of the ligand atom positions. The second set of features are contact distances between ligand and protein, which are processed by principal component analysis (PCA)\cite{Jackson1991,Sittel2018}. Both approaches permit the calculation of Euclidean distances in relevant feature spaces.

\paragraph{Ligand RMSDs} between trajectories $i$ and $j$ are calculated in the subspace of ligand positions as 
\begin{align}\label{eq:RMSD}
    d'^{\rm RMSD}_{ij}(t) = \sqrt{\frac{1}{L} \sum_{l=1}^L \lVert \mathbf{r}_{jl}(t) - \mathbf{r}_{il}(t) \rVert^2 }
\end{align}
where the sum runs over the $L$ ligand atoms $l$, and $\mathbf{r}_{il}$ denotes the Cartesian position of atom $l$ in trajectory $i$. 
To correct for drift from protein translational and rotational diffusion, we fit the protein backbone atoms to a reference structure for each time $t$. The same translation and rotation matrix adjusts the ligand accordingly.

When ligands diffuse freely within bulk water in the unbound state at large $x$ and $t$, naturally, larger distances occur. To counteract this, we suggested normalizing the distance\cite{Bray2022} with the mean over $N$ trajectories $\mean{d'^{\rm RMSD}(t)}_{N}$ evaluated at time $t$ as
\begin{align}\label{eq:RMSD_norm}
    d^{\rm RMSD}_{ij}(t) = \frac{d'^{\rm RMSD}_{ij}(t)}{\mean{d'^{\rm RMSD}(t)}_{N}}.
\end{align}

\paragraph{Ligand-protein contact PCA} is derived from minimal ligand-protein contact distances.\cite{Ernst2015} \SW{A preselection of contacts as input features is made from all protein residues $u$: a residue is considered a contact if its C\textsubscript{$\alpha$} atom is within a distance of \SI{0.45}{nm} from any atom of the ligand at least at one point in time within the trajectory ensemble.} Minimal distances $q_{i,u}(t)$ between the ligand and the heavy atoms of residue $u$ are recorded over time $t$ for each trajectory $i$ and referred to as contact distances. To reduce their dimensionality, we perform a PCA by calculating the covariance matrix
$\sigma_{uw} = \mean{ \left( q_u(t) - \mean{ q_u }_{N,t}\right) \left( q_w(t) - \mean{ q_w }_{N,t}\right) }_{N,t}$
averaging over $N$ trajectories and time $t$.\cite{Post2019}

Diagonalization of $(\sigma_{uw})$ yields eigenvectors $\{\mathbf{e}_k\}$, which are arranged in decreasing order according to their eigenvalues $\{\lambda_k\}$. The input coordinates of each trajectory $i$, the contact distances $\mathbf{q}_i(t) = (q_{i,1}(t), q_{i,2}(t), \dots)^\top$, are then projected onto the eigenvectors  via
$\mathrm{PC}_{ik}(t) = \mathbf{q}_i(t) \cdot \mathbf{e}_k $
to obtain the principal components (PCs). As a PCA is a unitary transformation, lengths are preserved by the projection. Furthermore, the contributions of contacts to the eigenvectors $\{\mathbf{e}_k\}$ allow for determining the impact of the individual input features and hence investigating the microscopic discriminants of the unbinding process.

Finally, a distance comparing the trajectories $i$ and $j$ is computed as
\begin{align}
    d^\text{PCA}_{ij}(t) = \sqrt{\sum_{k \in \Lambda} (\mathrm{PC}_{ik}(t) - \mathrm{PC}_{jk}(t))^2}.
\end{align}
In order to ensure a distance measure along relevant eigenvectors, remove noise from irrelevant dynamics, and facilitate visualization, usually only the first few $n$ PCs with dominant eigenvalues are considered\cite{Sittel2018}. They are contained in the set $\Lambda=\{1, 2, \dots, n\}$. In contrast to the RMSD approach, no normalization in the sense of Eq.~\eqref{eq:RMSD_norm} is performed.

\subsubsection{Similarity}
Both types of input features described above result in a time-dependent distance $d_{ij}(t)$, which is reduced by computing the time average $d_{ij}=\mean{d_{ij}(t)}_t$. From this dissimilarity between two trajectories $i$ and $j$, a similarity matrix $(s_{ij})$ is calculated as
\begin{equation}\label{eq:similarity_from_distance}
    s_{ij} = 1-\frac{d_{ij}}{d_\mathrm{max}},
\end{equation}
where $d_\mathrm{max}$ corresponds to the largest distance $d_{ij}$.

\subsubsection{Clustering via Leiden community detection}
To group trajectories according to their similarity $s_{ij}$, we employ the Leiden community detection algorithm\cite{Traag2011,Traag2019,Diez2022}. This graph-based method allows representing trajectories as nodes and encoding the similarities between the trajectories $s_{ij}$ as edges. The Leiden algorithm identifies clusters by maximizing an objective function, for which we employ the Constant Potts Model (CPM)
\begin{equation}\label{eq:CPM}
    \Phi_\mathrm{CPM} = \sum_{c} \left( e_{c} - \gamma \binom{n_{c}}{2} \right).
\end{equation}
Here, $e_{c}$ denotes the sum of all similarities within a cluster $c$ and $n_{c}$ represents the number of trajectories in $c$, resulting in $\binom{n_{c}}{2}$ similarities in $c$. This number is scaled by the user-defined resolution parameter $\gamma$. Thus, $\gamma \binom{n_{c}}{2}$ describes a hypothetical cluster of the same size as $c$, but with constant similarities $\gamma$. The aim of the maximization of Eq.~\eqref{eq:CPM} is to find clusters whose summed similarities $e_c$ surpass the ones from their hypothetical counterparts. Therefore, $\gamma$ controls the coarseness of the clustering, since trajectories in a cluster have to be at least similar with a value of $\gamma$ on average. This allows for a simple separation of outliers. Notably, some similarities in a cluster may be below $\gamma$, if the objective function overall benefits from it. Hence, $\gamma$ is not a cutoff in the sense of hierarchical clustering\cite{Bray2022}.

To illustrate results, we permute $(s_{ij})$ into a block matrix, as exemplified in Fig.~\ref{fig:Stb_Panel}\,B. The individual blocks represent the clusters following the same pathway with respect to the used input feature.

\section{Methods}

\subsection{Streptavidin-biotin complex}

Streptavidin-biotin unbinding trajectories were reported on in Ref.\citenum{Cai2023}. In brief, a St-b tetramer was generated based on PDB ID 3RY2\cite{LeTrong2011} and its dynamics simulated employing the Amber99SB*-ILDN force field\cite{Hornak2006,Best2009,Lindorfflarsen2010}. The tetramer was solvated with TIP3P water molecules\cite{Jorgensen1983} and oriented in such a way that the unbinding path of biotin within one of the monomers is parallel to the $x$ axis of the simulation box. This St-b monomer is later used for pulling simulations. Position restraints were applied on selected C$_\alpha$ atoms in said monomer to keep its orientation stable. Biotin force field parameters were generated employing antechamber\cite{Wang2006}, acpype\cite{SousadaSilva12} and GAFF\cite{Wang2004} atom parameters. Biotin atomic charges were calculated using ab initio calculations on the HF/6-31G* level in Orca\cite{Neese2012} followed by RESP charge fitting in Multiwfn\cite{Lu2012a}. Simulations were run using the PULL code of Gromacs\cite{Abraham15} v2016.4 and v2020.4 with a constant constraint velocity of $0.1\,\text{m}/\text{s}$ along a defined pulling vector in Cartesian space, where the ligands can diffuse on the moving plane orthogonal to the pulling vector, over a distance of $2\,\text{nm}$. The used pulling vectors are defined in a $(x, y, z)$ nomenclature: (1, 0, 0), (1, 1, 0), (1, -1, 0), (1, 0, 1) and (1, 0, -1) with the following number of trajectories per direction: (1, 0, 0): 200,
(1, 0, 1): 71, (1, 0, -1): 87, (1, 1, 0): 93, (1, -1, 0): 142. The Cartesian coordinate $x$ is not to be confused with the biasing coordinate.

\subsection{A\down{2a} adenosine receptor}

A\down{2a} adenosine receptor simulations with bound antagonist ZM241385 are based on the crystal structure of a thermostabilized receptor mutant with PDB ID 5IU8\cite{Segala2016}. We explicitly kept the thermostabilizing mutations within the heptahelical transmembrane motif, as they restrict protein conformational dynamics, which facilitates pathway detection. The intracellular loop 3 was regenerated using SWISS-MODEL\cite{Waterhouse2018}. Side chain orientations and protonation states were determined using PROPKA3\cite{Olsson2011}. We encountered protein instabilities with a neutrally charged His278 in helix 7. Due to its proximity to Glu13 in helix 1, doubly protonating and thus positively charging His278 established a salt bridge between the two residues, which resolved the instability. Protein-internal water molecules as well as the central sodium ion present in the crystal structure were kept. The protein was embedded in a pre-equilibrated membrane model of 128 POPC lipids\cite{Berger1997} surrounded by a $0.154\,\text{M}$ NaCl solution with TIP3P water molecules employing the INFLATEGRO script\cite{Schmidt2012}.
The system was described using a combination\cite{Cordomi2012} of Amber99SB force field parameters\cite{Hornak2006,Best2009} for the protein and Berger lipid parameters\cite{Berger1997,Tieleman1999}. Ligand parameters were derived by the same procedure as for biotin described above. Equilibration followed a procedure described in Refs.~\citenum{Jaeger2021} and \citenum{Wolf2008b} with a short final equilibration of $20\,$ns length. Simulations were carried out with Gromacs versions as noted above using an $1\,\text{fs}$ integration step, the Bussi v-rescale thermostat\cite{Bussi2007} ($\tau_T = 0.2\,\text{ps}$) at a temperature of $290.15\,\text{K}$, and the Parrinello-Rahman barostat\cite{Parrinello1981} ($\tau_P = 0.5\,\text{ps}$) at $1\,\text{bar}$ with semi-isotropic pressure coupling for equilibration and isotropic coupling for production runs. After equilibration, the simulation box was quadruplicated to calculate four unbinding events within a single simulation, which allowed longer pulling distances of up to $3.5\,\text{nm}$ due to the enlarged simulation box. Pulling simulations were initiated by generating 200 simulation run inputs from the equilibrated four-protein system with independent velocity distributions, followed by a short $20\,\text{ps}$ equilibration with position restraints on proteins and ligand. Pulling simulations were then calculated with the Gromacs PULL code at a pulling velocity of $0.1\,\text{m}/\text{s}$ over a distance of $3.5\,\text{nm}$. As biasing coordinate, the distance between the centers of mass of the ligand's heavy atoms and a set of backbone atoms within the middle of the transmembrane helices (residues No. 14, 58, 59, 83, 84, 85, 136, 182, 183, 246, 247, 276, 277 and 278, with a center of mass close to the central sodium ion) is used per protein, resulting in 804 unbinding events in total. We discarded 123 of the original unbinding events due to direct contacts between ligands of adjacent protein monomers during simulation, leaving us with 681 trajectories to analyze.

\subsection{Data evaluation and visualization}

Leiden clustering was performed with the MoSAIC tool\cite{Diez2022}. To benchmark the Leiden clustering 
results, we employ \textit{k}-medoids clustering\cite{Madhulatha2012}, also implemented in MoSAIC, for which the Silhouette score (see the SI) can guide the optimal number of medoids. Clustering results were additionally compared based on homogeneity scores\cite{Rosenberg07} (see the SI).
Ligand-protein minimal distances and ligand RMSDs, using the protein backbone as fitting group, were calculated with MDAnalysis\cite{mdanalysis2011, mdanalysis2016}. 
Sankey plots were generated with pySankey\cite{pysankey}. 
\SW{Path-wise free energy profiles were calculated according to Eq.~(\ref{eg:dcTMD_dGpathwise}).}
Molecular structures and volume maps were visualized using VMD\cite{Humphrey1996} unless otherwise stated.

\section{Results and discussion}

\subsection{Benchmarking with St-b}

\begin{figure}[h!]
    \centering
    \includegraphics[width=0.45\textwidth]{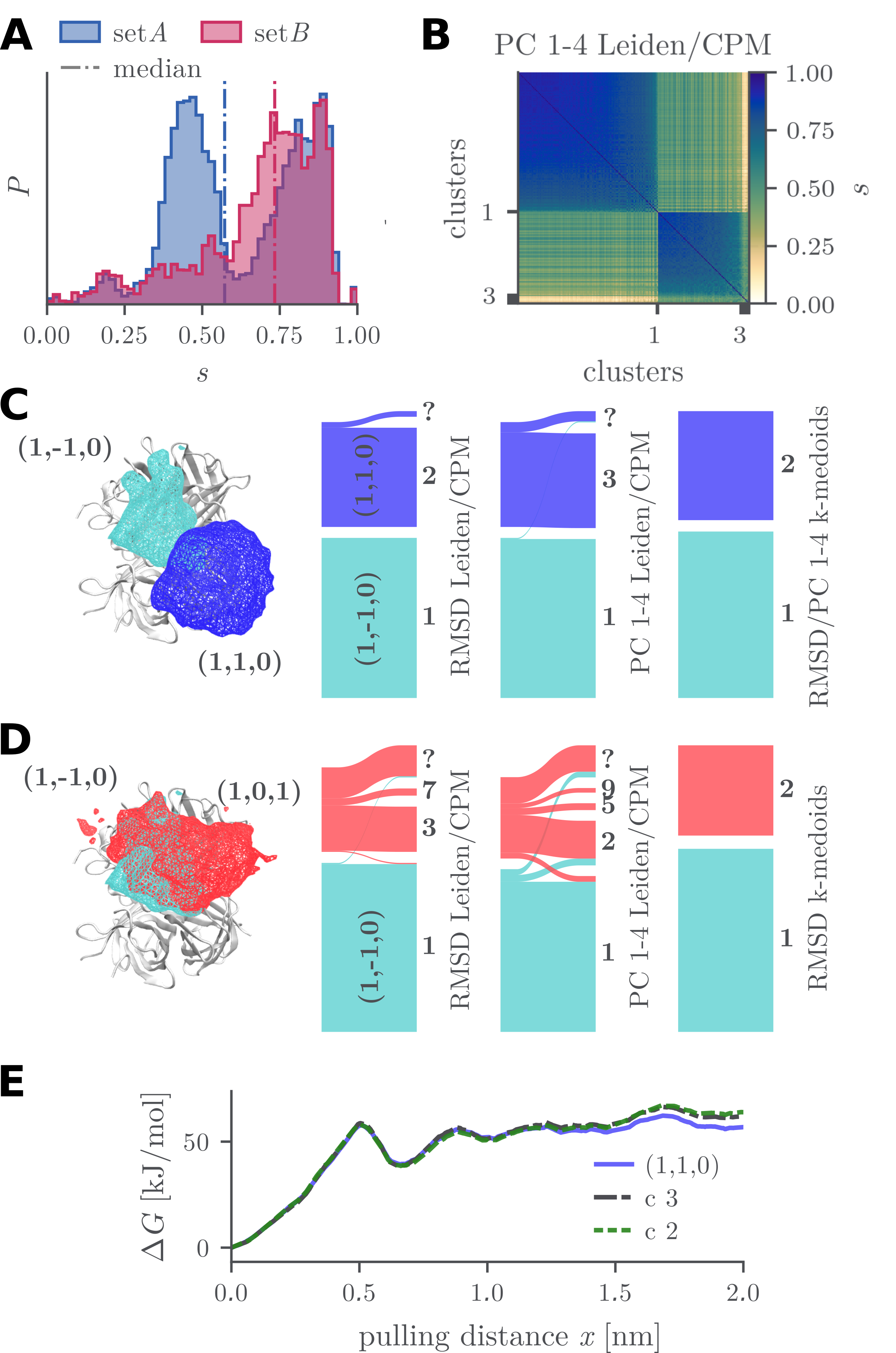}
    \caption{Streptavidin-biotin pathway identification benchmarking. A: distribution of \SW{conPCA}-based similarities $s_{ij}$. B: similarity block matrix using \SW{conPCA}-based input and Leiden clustering on subset $A$. C: visualization of set $A$ and clustering results as Sankey plots. D: visualization of set $B$ and clustering results as Sankey plots. Volume maps of biotin positions as wire frames. Visualizing the trajectories pulled along the vectors: $(1, -1, 0)$ in cyan, $(1, 1, 0)$ in blue and $(1, 0, 1)$ in red. Clusters with $N<3$ are combined into a single macro-cluster labeled with ``?''. \SW{E: comparison of estimated free energies $\Delta G$ for $(1, 1, 0)$ ground truth (blue) and the corresponding RMSD (green dashed) and conPCA-based (grey dashed) clusters 2 and 3, respectively.}}
    \label{fig:Stb_Panel}
\end{figure}
The St-b trajectory data from our earlier study\cite{Cai2023} is an ideal benchmark system for trajectory clustering, as the protein does not display significant conformational dynamics during enforced unbinding
and the pathways in Cartesian space are known (see Fig.~\SIStbsystem\ for a structural overview).

We generated two test sets $A$ and $B$ with each set combining trajectories from two different pulling directions: set $A$ consists of trajectories pulled along the vectors $(1, 1, 0)$ and $(1, -1, 0)$, while set $B$ consists of trajectories pulled along $(1, 0, 1)$ and $(1, -1, 0)$. Visualizing density isosurfaces (``volume maps'') of the ligand trajectories from these subsets in Fig.~\ref{fig:Stb_Panel}\,C,\,D reveals that subset $A$ comprises well-separated trajectories, whereas subset $B$ shows a pronounced overlap between the trajectories. This is reflected in the similarity distributions in Fig.~\ref{fig:Stb_Panel}\,A, where set $B$ has a higher median. The ground truth, i.e., the two pulling directions, should be easily recoverable for set $A$, while set $B$ should pose a challenge for any separation approach.
To add another layer of difficulty, we generated a third set $C$ in which we combined all three subsets, i.e., $(1, 1, 0)$, $(1, 0, 1)$ and $(1, -1, 0)$.

First, for all sets, the RMSD and conPCA-based similarity matrices $(s_{ij})$ were calculated. For the latter, we tested principal components 1+2, 2+3, 1-3 and 1-4 and identified PCs 1-4 as a suitable subspace which contains most of the relevant information. The absolute difference between the similarities obtained from the full space and the ones from PCs 1-4 is $<0.1$, as Fig.~\SIStbsim\ displays. \SW{The information required for trajectory clustering is therefore already contained in the first four PCs, which we use for calculating similarities.}

We proceeded with the determination of a suitable resolution factor $\gamma$. We tested values between $0.4$ and $0.9$. \SW{For the 681 trajectories, each clustering takes only a few seconds.} Comparing the clustering results in Figs.~\ref{fig:Stb_Panel}\,C,\,D, \SIStbsimRMSD\ and \SIStbsimPCA\, we find that a $\gamma\approx\text{median}(s_{ij})$ recovers the ground truth, i.e., the subsets of $A$ and $B$, best. This is reflected in homogeneity scores of $h= 0.98$ for set $A$ and $h= 0.62$ for set $B$ when using PCs 1-4, and $h=1$ for set $A$ and $h=0.92$ for set $B$ using RMSD based input features. 

\SW{To assess the quality of free energy predictions given by the clustering, we compare the ground truth free energy of the  $(1, 1, 0)$ trajectories from set $A$ to the corresponding clusters from RMSD- and conPCA-based similarities. As Fig.~1E shows, the differences between the ground truth and the clustering results are negligible.}
In general, lower resolution parameters $\gamma$, here  $\gamma < \text{median}(s_{ij})$, result in significant mixing of clusters in sets $B$ and $C$. Using a higher $\gamma$ enforces more homogeneous clusters and thus the identification of various smaller clusters in all investigated sets.
\SW{Similarily, as Figs.~S5 and S6 show, a low $\gamma$ yields few clusters with a large number of dissimilar trajectories, whereas a too high $\gamma$ results in numerous small clusters with high inter-cluster similarities.}
We therefore recommend using $\gamma = \text{median}(s_{ij})$ as starting point for pathway identification. However, it is important to note that this reasoning may be affected by the fact that our benchmark sets $A$ and $B$ are composed of only two subsets. This is reflected in set $C$, where a higher $\gamma$, between the median and the third quartile of $(s_{ij})$, yields the best clustering results.

Exploring \textit{k}-medoids clustering as an alternative approach, Figs.~\ref{fig:Stb_Panel}\,C,\,D, \SIStbRMSDsankey\ and 
\SIStbPCAsankey\ show that this method outperforms the Leiden community detection for all three investigated sets. This follows from the design of \textit{k}-medoids, as it requires that the number of medoids, and therefore clusters, is known a priori. An appropriate number of clusters can be estimated, e.g., via the Silhouette score. Figs. \SIStbRMSDsilhouette, \SIStbPCAsilhouette\ display such an analysis for all three sets and both similarity measures employed: for the benchmark sets $A$ and $B$, the Silhouette score is indeed maximal for the number of pulling directions within the sets $A$ and $B$, but not for set $C$ when using conPCA-based similarities. When joining all five pulling directions simulated by us earlier, \textit{k}-medoids fails for both RMSD- and conPCA-based similarities.

In summary, \textit{k}-medoids clustering demonstrates superior performance over Leiden community detection in our benchmark set. However, its effectiveness is contingent on knowing $k$, necessitating caution in scenarios beyond our benchmark system. For Leiden clustering, $\gamma = \text{median}(s_{ij})$ is a good starting point for pathway identification. In contrast to \textit{k}-medoids clustering, outlier trajectories are separated without needing a user to know how many outliers there are. Clusters based on conPCA and this resolution parameter are more numerous and smaller than those based on RMSD, but the resulting eigenvectors provide insights into the impact of specific residues on ligand pathways and thus offer valuable information on the underlying microscopic mechanisms\cite{Wolf2020}.

\subsection{A\down{2a} ligand unbinding}
\begin{figure}[htb!]
	\centering
	\includegraphics[width=\linewidth]{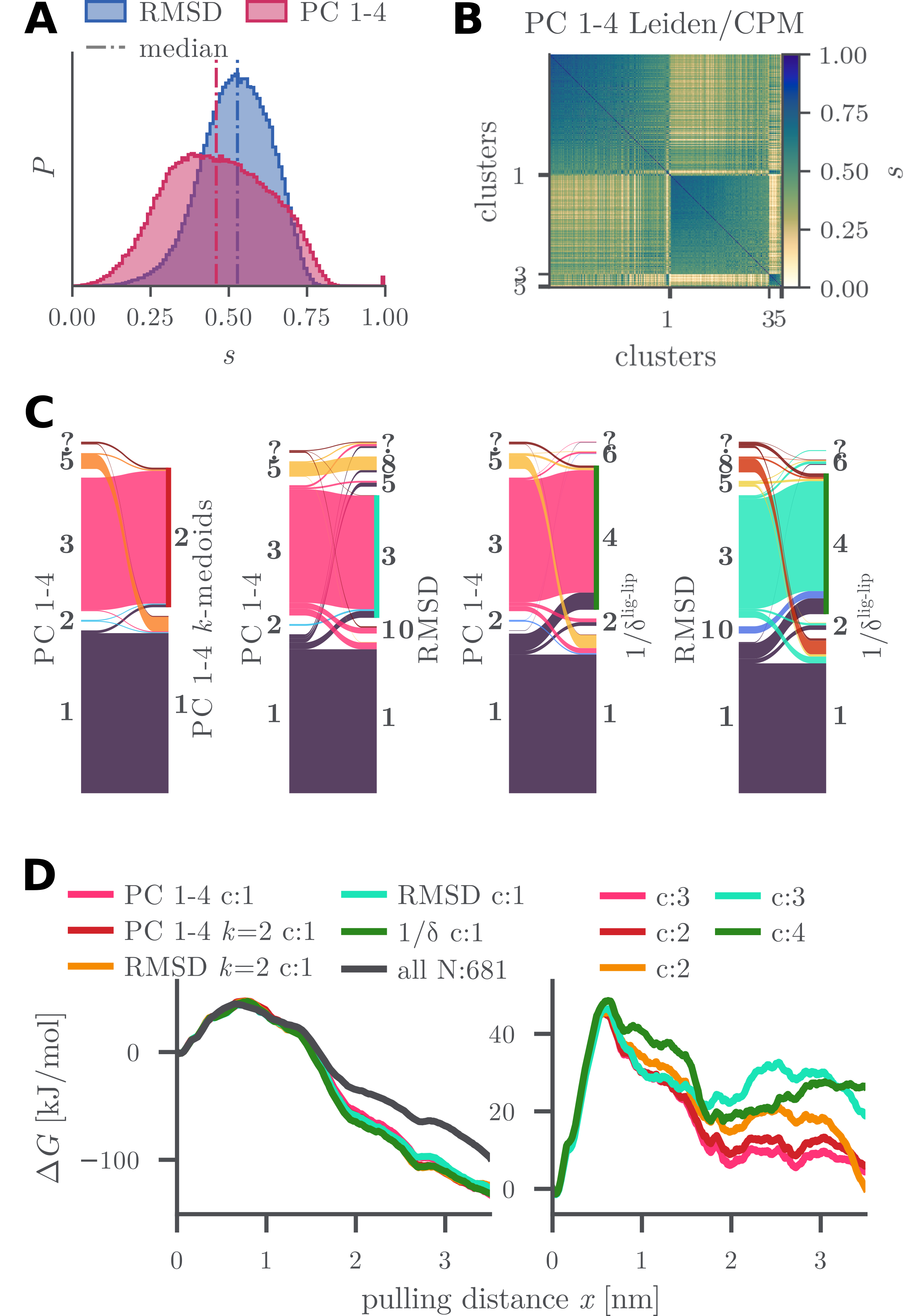}
\caption{Trajectory clustering for pathway identification in A\down{2a}. A: RMSD and conPCA based similarity distribution. B: similarity block matrix based on PCs 1-4 at $\gamma = 0.46$ (equal to $\text{median}(s_{ij})$). C: Sankey plots compare the cluster contents of clustering results based on the input features PCs 1-4, RMSD and inverse ligand-lipid minimal distances  $1/\delta^\text{lig-lip}$. Clusters with $N<3$ are combined into a single macro-cluster labeled with ``?''. D: comparison of estimated free energies $\Delta G_k$ of clusters containing $N>100$ trajectories using the input features PC 1-4, RMSD and $1/\delta^\text{lig-lip}$. ``c'' indicates cluster number, ``k'' indicates the number of medoids (if applicable). The line color signifies the input feature type and clustering method. Left: clusters with $\Delta G_k<\SI{-10}{kJ/mol}$. Right: clusters that solve the friction overestimation.
}
\label{fig:A2A_panel}
\end{figure}
Following our benchmark with St-b, we applied our workflow to a more challenging protein-ligand complex, a thermostabilized variant of the A\down{2a} receptor\cite{Segala2016} bound to the antagonist ZM241385 (see Fig.~\SIAdeSystem\ for an overview of the simulation system).
\SW{Consulting the similarity distributions in Fig.~\SIAdesim, we find again that the first four PCs contain the information relevant for trajectory clustering.} On the basis of the good clustering results in our benchmark system, we utilized $\gamma \approx \text{median}(s_{ij})$.
The resulting similarity block matrices in Figs.~\ref{fig:A2A_panel}\,B and \SIAdeRMSDsimblock\ show two major clusters each. Performing dcTMD analysis as displayed in Fig.~\ref{fig:A2A_panel}\,D and Fig.~\SIAdefel\ shows that in each of the similarity block matrices only one cluster exhibits a mitigated friction overestimation. The other clusters exhibit artificially low free energies, due to a remaining non-normal work distribution, hinting at additional hidden CVs that we cannot resolve with our input features. As Fig.~\ref{fig:A2A_panel}\,C indicates, both RMSD and conPCA similarities result in comparable clusters with negligible mixing within the Sankey plots. Consequently, as can be seen in 
Fig.~\ref{fig:A2A_panel}\,D, they both result in free energy curves with a barrier height of $\Delta G^{\neq}\approx 45\,$kJ/mol. The free energy in the unbound state is $\approx 5\,$kJ/mol for PCs 1-4-based clusters and $\approx 25\,$kJ/mol for RMSD clusters.

To assess the robustness of the predicted path-wise free energies, we tested the impact of a  resolution factors $\gamma$, up to the third quartile of $(s_{ij})$, on the clustering results of both input features. Fig.~\SIAdefel\ displays the resulting free energy estimate: with the third quartile, the number of trajectories contained in a cluster declines drastically (by about 40\%). Despite this, the barriers $\Delta G^{\neq}$ show only a moderate increase of $\approx5$-$10\,\text{kJ}/\text{mol}$. In the unbound state, however, the free energies increase to $\approx25\,$kJ/mol for conPCA-based clusters and up to $\approx50\,$kJ/mol for RMSD clusters.

This observation matches the bootstrapping errors for the Leiden clustering results using $\gamma \approx \text{median}(s_{ij})$ (see Fig.~{\SIAdeLeidenfelbootstrapp}). While the error associated with the barrier is negligible, it reaches up to $\pm\,\SI{20}{kJ/mol}$ in the unbound state.
A closer analysis of the cluster composition for increasing $\gamma$ shows, comparable to the results for St-b, that the main clusters are mostly split into subclusters, yet, there is no significant mixing (see Fig.~{\SIAdeLeidensankey}).

In order to evaluate the capabilities of Leiden/CPM clustering, we compare it with other possible clustering methods, namely Leiden/Modularity\cite{Newman2004}, complete linkage \cite{Sorensen1948}, $k$-medoids, and NeighborNet\cite{Huson2006}, as used by us before\cite{Bray2022}. The results of these tests are presented in Figs.~\SIAdekmedsilhouette\ to {\SIAdeNeighbor}. They show that the former two approaches are outperformed by Leiden/CPM, whereas $k$-medoids performs comparably. NeighborNet spreads the satisfactory RMSD cluster from Leiden/CPM across several clusters. 

In summary, Leiden/CPM proves to be a well-performing trajectory clustering tool with minimal user input which mitigates the friction overestimation artifact, and results in reasonably robust free energy barriers for different values of $\gamma$. The main remaining issue is that this mitigation is only given for a single trajectory cluster. In general, finding the origin of the occurrence of pathways is limited by the chosen input features. With the information gained by identifying clusters, we reconsider our system in search for a better input feature and thus an improved RC candidate.

\subsubsection*{Exploring microscopic pathway origins}
\begin{figure}[htb!]
    \centering
    \includegraphics[width=\linewidth]{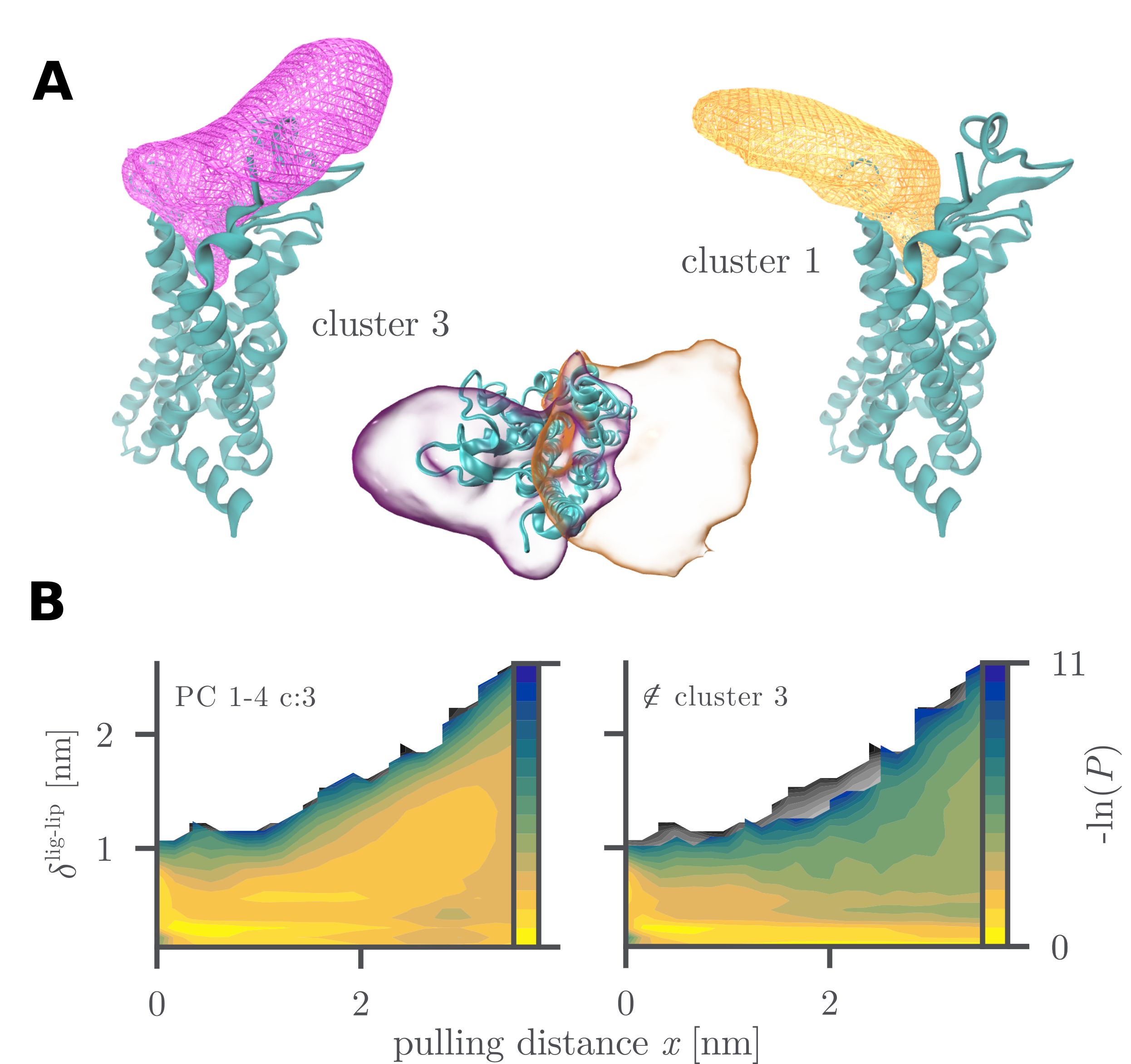}
    \caption{A: volume accessed by ZM241385 in clusters 1 and 3, respectively, based on PCs 1-4 as displayed in Fig.~\ref{fig:A2A_panel}. B: probability distribution of the ligand-lipid minimal distance $\delta^\text{lig-lip}$ over the pulling distance $x$ of all trajectories in grayscale, overlaid by the distribution of cluster 3  (left) and its complement (right) in color.}
    \label{fig:A2A_liglipPanel}
\end{figure}
To gain a better microscopic understanding of the ligand's unbinding path, we visualized the volume accessed by the trajectories contained in the two main conPCA clusters 1 and 3 from Fig.~\ref{fig:A2A_panel}, of which cluster 3 resolved the friction overestimation artifact. As Fig.~\ref{fig:A2A_liglipPanel} shows, \SW{both clusters take the same route from the binding site to the extracellular surface of the receptor. Here, the clusters bifurcate: cluster 3 moves along the extracellular loop 2 (EL2) of the receptor, whereas cluster 1 passes through the cleft between transmembrane helices 1 and 7. Since this latter route takes the ligand close to the membrane lipids, ZM241385-lipid contacts may influence the emergence of pathways.} Indeed, considering the pulling distance resolved distributions of the minimal ligand-lipid distances $\delta^\text{lig-lip}(x)$ in Fig.~\ref{fig:A2A_liglipPanel}, we find that ligands in cluster 3 move away from the membrane after a pulling distance of $\approx 1\,$nm, whereas ligands that do not belong to cluster 3 become bound to the membrane. This observation is in line with a rather high predicted hydrophobicity index for ZM241385 ($\log{P} = 2.93$).\cite{ZMAChEMBL}

The characterization of friction overestimation resolving clusters in both RMSD- and conPCA-based clustering becomes more evident when plotting inverse minimal distances $1/\delta^\text{lig-lip}$ as shown in Fig.~\SIAdeinvlipid.
This inspired the use of inverse ligand-lipid distance time traces $1/\delta^\text{lig-lip}_i(t)$ as input features.

We repeat our clustering approach using Leiden. 
Figs.~\ref{fig:A2A_panel}\,D,\,E and \SIAdeAllsankey\ show cluster compositions and free energy profiles comparable to the ones gained from RMSD or conPCA. Thus, we assume that the friction overestimation of cluster 1 comes from ligands entering the membrane at random times and being pushed through the water-membrane interface. We have shown earlier that dcTMD is not suited for such clusters with random switching between different regimes of dynamics.\cite{Wolf2023}

Focusing on the unbinding trajectory clusters from the point of view of A\down{2a}-ligand contacts, the route of cluster 3 moves along EL2, with which the ligands form contacts (see Fig.~\SIAdecontacts). This unbinding path is in good agreement with the ZM241385 unbinding behavior described in other studies.\cite{Mattedi2019,Deganutti2020} Additionally, it has been observed for the $\beta_2$ adrenergic receptor that EL2 forms a temporary ligand position for ligand desolvatisation,\cite{Dror2011a} which may be the case here as well, though we do not see an accompanying minimum in our free energy profiles of A2\down{a}.

\section{Conclusions}
In this study, we have introduced a straightforward, easily applicable and customizable machine learning approach for the identification of unbinding pathways from a set of biased protein-ligand dissociation trajectories. Built on community detection, the method only requires the definition of a suitable similarity measure and a resolution parameter.
\SW{We find $\gamma \approx \text{median}(s_{ij})$ to be a good empirical value as a starting point for the search for an optimal resolution parameter.} In our benchmark with the streptavidin-biotin complex, RMSD-based similarities slightly outperform their conPCA-based counterparts, while they perform similarly well for the A\down{2a} adenosine receptor with inhibitor ZM241385. 
In this challenging system, our approach succeeds in finding a relevant pathway, for which we present estimated free energy profiles. On top of that, conPCA-based similarities hold the potential to offer insight into the microscopic dynamics and lead the way towards meaningful RC candidates. The input feature space is reduced from many contact distances to four PCs and finally one distance between the ligand and the lipid bilayer as meaningful RC. \SW{An optimal RC however is highly dependent on the studied biomolecular complex, and needs to be revealed individually for each protein-ligand complex of interest\cite{Bray2022,Wolf2023a}.}

We note that while the predicted A\down{2a} receptor complex free energy barriers are robust, free energy estimates of the unbound state still vary strongly. In upcoming studies, we will investigate under which conditions the estimated profiles can be used for the prediction of unbinding and binding kinetics via Langevin equation simulations\cite{Wolf2020} in comparison to experimentally determined kinetics.

While our approach has been tailored to the needs of finding pathways in sets of trajectories with a constant velocity constraint, finding pathways in dynamical data is a far more general topic\cite{Yuan2017} with different approaches needing different distance measures. Biased unbinding simulations employing steered MD or infrequent metadynamics require a similarity measure taking into account unbinding events at different times. Worth mentioning here is a recent study that employs dynamic time warping\cite{Ray2023a} as a distance measure. \SW{Employing other distance measures such as Hausdorff or Fr\'echet distances\cite{Yuan2017} may be of further interest.} Our classification learning via Leiden/CPM as implemented in MoSAIC is flexible to work with any distance and similarity measure available. \SW{Similarly, our approach should be capable to classify trajectories of ligand binding, as well.}

\section{Data and software availability}

The dcTMD analysis tool as well as the Leiden clustering tool MoSAIC are available at \url{https://www.moldyn.uni-freiburg.de/software.html}. St-b and A\down{2a}-ZM241385 unbinding trajectories are available from the authors upon reasonable request.

\begin{acknowledgement}

This work has been supported by the Deutsche
Forschungsgemeinschaft (DFG) via grant WO 2410/2-1 
within the framework of the Research Unit FOR~5099 
``Reducing complexity of nonequilibrium'' (project No.~431945604), 
the High Performance and Cloud Computing Group at the Zentrum f\"ur Datenverarbeitung of the
University of T\"ubingen, the state of Baden-W\"urttemberg through bwHPC and the DFG through grant no INST 37/935-1 FUGG (RV bw16I016). The authors are grateful to Gerhard Stock, Georg Diez, Fabian Koch, Daniel Nagel, Matthias Post, and Anja Seegebrecht (all University of Freiburg) for helpful discussions.

\end{acknowledgement}

\providecommand{\latin}[1]{#1}
\makeatletter
\providecommand{\doi}
  {\begingroup\let\do\@makeother\dospecials
  \catcode`\{=1 \catcode`\}=2 \doi@aux}
\providecommand{\doi@aux}[1]{\endgroup\texttt{#1}}
\makeatother
\providecommand*\mcitethebibliography{\thebibliography}
\csname @ifundefined\endcsname{endmcitethebibliography}
  {\let\endmcitethebibliography\endthebibliography}{}

%%%%%%%%%%%%%%%%%%%%%%%%%%%%%%%%%%%%%%%%% SI
%%%%%%%%%%%%%%%%%%%%%%%%
%%%%%%%%%%%%%%%%%%%%%%%%

\clearpage
\onecolumn

\renewcommand{\thepage}{\arabic{chapter}.\arabic{page}}  
\renewcommand{\thesection}{\arabic{chapter}.\arabic{section}}   
\renewcommand{\thetable}{\arabic{chapter}.\arabic{table}}   
\renewcommand{\thefigure}{\arabic{chapter}.\arabic{figure}}
\renewcommand{\thepage}{S\arabic{page}}  
\renewcommand{\thesection}{S\arabic{section}}   
\renewcommand{\thetable}{S\arabic{table}}   
\renewcommand{\thefigure}{S\arabic{figure}}

\setcounter{page}{1}    
\setcounter{section}{0}    
\setcounter{figure}{0}    
\setcounter{enumiv}{0}

%%%%%%%%%%%%%%%%%%%%%%%%%%%%%%%%%%%%%%%%%%
%%%%%%%%%%%%%%%%%%%%%%%% SI %%%%%%%%%%%%%%%%%%
%%%%%%%%%%%%%%%%%%%%%%%%%%%%%%%%%%%%%%%%%%

\section*{Supporting Information}

\section{Silhouette score}\label{chap:silhouette_theory}
%\todo[inline]{We have a problem here, as we already extensively use $s$ for the similarity. What do we do, use a different symbol for the Silhouette? I think it is fine, since we never use s to indicate the silhouette score in the text but always write silhouette score. I used a capital s for the mean Silhouette score. i think there will be no confusionwith the rest...}
The Silhouette score allows to evaluate the quality of a clustering depending on the used input parameters such as \textit{k}, the number of medoids in \textit{k}-medoids, or the resolution factor $\gamma$. In our case, the score is calculated per trajectory $i$. First, the mean distance $a_i$ in the cluster to which trajectory $i$ belongs, denoted as $c_I$, is defined as 
\begin{align}
    a_i &= \frac{1}{\left| c_I \right| - 1} \sum_{c_I \ni j \neq i} d(i, j)\\ 
    \intertext{where $\left| c_I \right|$ is the number of trajectories in the cluster $c_I$. The smaller $a_i$ is, the better the trajectory fits in its cluster. Then, the mean distance to the neighboring cluster is defined via}
    b_i &= \min_{J\neq I}\frac{1}{\left| c_J \right|} \sum_{j \in c_J} d(i, j),\\
    \intertext{which can also be expressed as the smallest mean distance to all trajectories of another cluster. The Silhouette score $s_i$ is then written as}
    s_i &= \begin{cases}
        \frac{b_i - a_i}{\max\left\{ a_i, b_i \right\}} & \text{if $\left|c_I\right|>0$}\\
        0 & \text{if $\left|c_I\right|=0$}
    \end{cases}
\end{align}
The mean Silhouette score $S=\left< s_i \right>_i$ can then be used to assess the quality of the clustering. An optimal $S$ is close to $1$, the worst possible value is $-1$. A maximum of $S$ therefore suggests an optimal parameter choice.

\section{Homogeneity score}
Clustering results are compared based on homogeneity scores\cite{Rosenberg07}
\begin{equation}
    h = \begin{cases} 1 & \text{if } H(G,K) = 0\\
         1-\frac{H(G|K)}{H(G)} & \text{else}\end{cases}
\end{equation}
where a data set with $N$ data points (in our case trajectories) is assumed with a set of ground truth classes $G=\{g_i|i=1, ...,n\}$ with $n\leq N$ and a set of clusters $K=\{k_i|1,\dots,m\}$ with $m \leq N$. $H(G|K)$ is the conditional Shannon entropy of the class distribution given the proposed clustering, which is normalized by the maximum reduction in entropy through the clustering entropy $H(G)$, and $H(G,K)$ is the respective joint entropy. In our application, the maximal $h=1$ is given if all found clusters $K$ each only contain trajectories from one ground truth trajectory set $G$, whereas mixed attributions lower $h$.

\section{Streptavidin-biotin tetramer}
\subsection{Structural overview}
\begin{figure}[H]
\centering
	\begin{subfigure}{0.4\linewidth}
	\includegraphics[width=\linewidth]{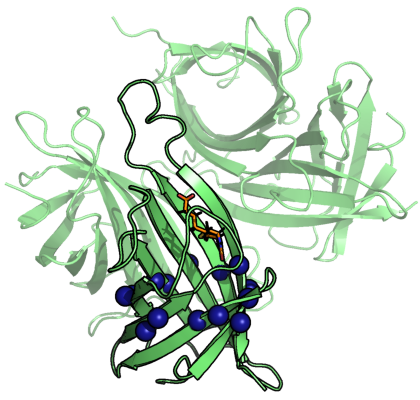}
    \caption{}
	\end{subfigure}
 	\begin{subfigure}{0.4\linewidth}
	    \includegraphics[width=\linewidth]{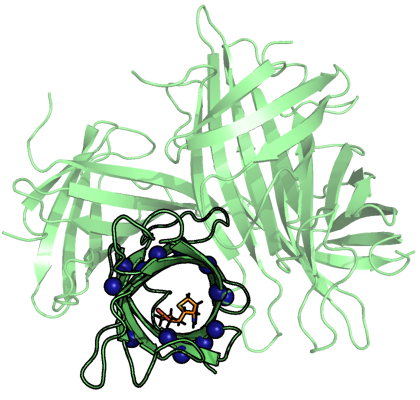}
	    \caption{}
	\end{subfigure}
 \caption{Streptavidin-biotin tetramer system in (a) side and (b) top view. The streptavidin (green cartoon) monomer used for pulling is in the foreground bound to biotin shown as orange sticks. The blue spheres indicate the  C$_\alpha$ atoms of the pull group.}
\end{figure}

\section{Importance of the PCs for conPCA similarities: St-b}
\begin{figure}[H]
	\centering
	\begin{subfigure}{0.7\linewidth}
	\includegraphics[width=\linewidth]{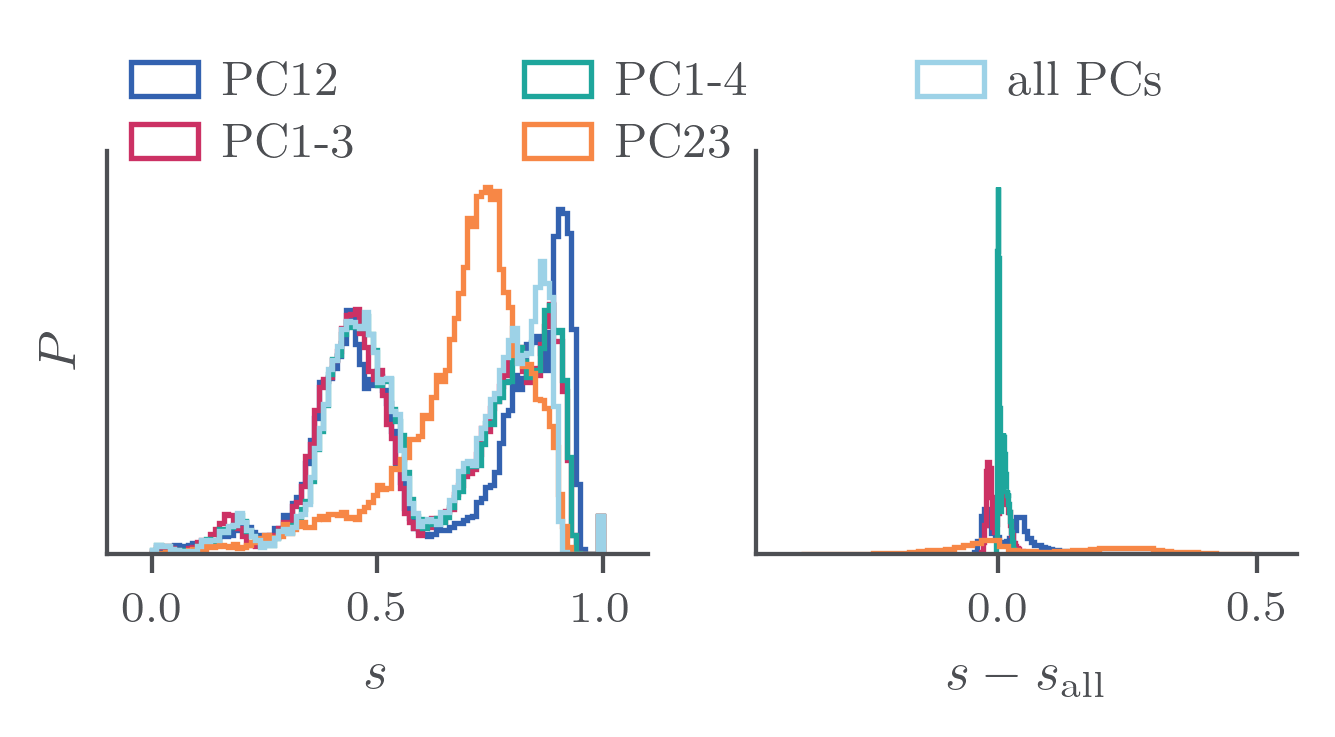}
    \caption{St-b subset A (pulling vectors (1, 1, 0) and (1, -1, 0))}
	\end{subfigure}
 	\begin{subfigure}{0.7\linewidth}
	       \includegraphics[width=\linewidth]{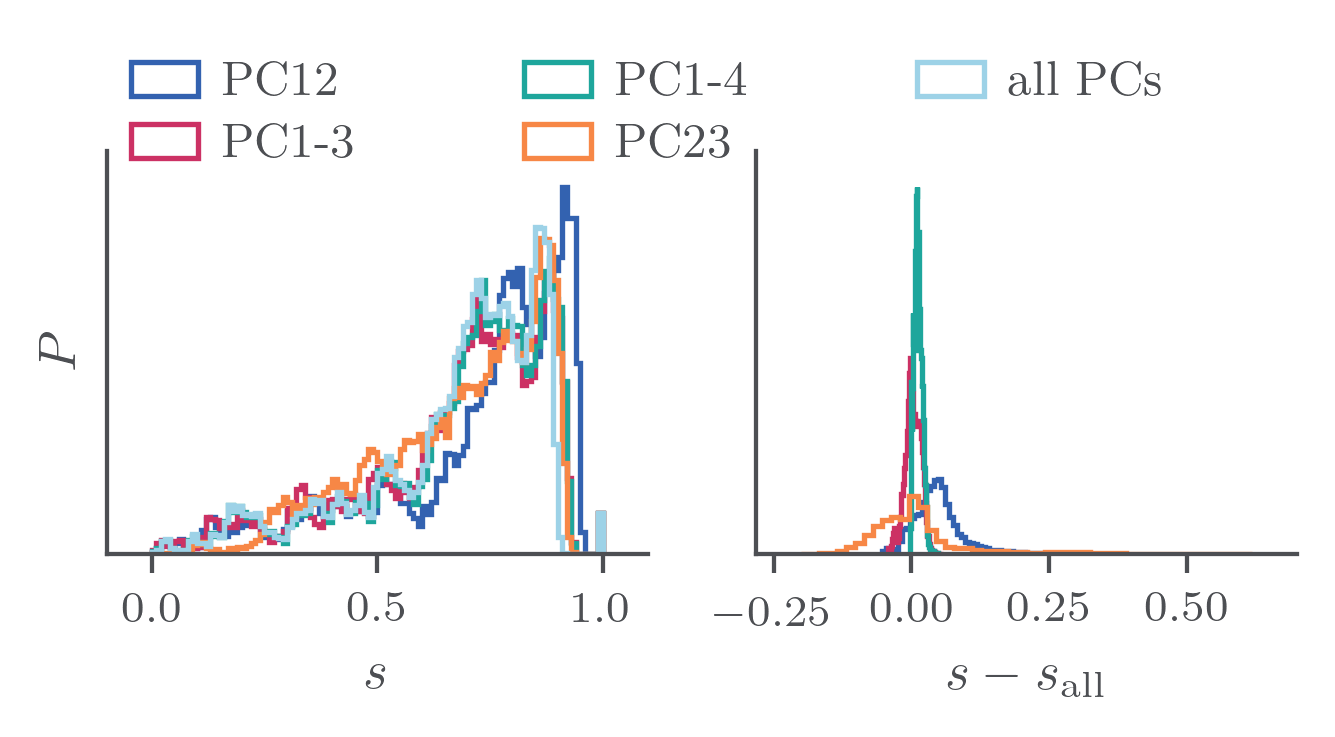}
	       \caption{St-b subset B (pulling vectors (1, 0, 1) and (1, -1, 0))}
	   \label{fig:St-b_xyz1-10xyz101_comparePCs_similarity_distribution}
	\end{subfigure}
 \caption{Left: probability distribution of the similarities $s$ calculated using different PCs. Right: probability distribution of the element-wise differences between similarities calculated based on PC subsets (e.g. PC12, PC1-3 etc.) and the full PCA space $s_\mathrm{all}$ (all PCs).}
	\label{fig:St-b_comparePCs_similarity_distribution}
\end{figure}

\section{Clustering on St-b}
\subsection{Similarity distributions}
\subsubsection{RMSD}
% TODO: \usepackage{graphicx} required
\begin{figure}[H]
	\centering
	\begin{subfigure}{0.3\linewidth}
		\includegraphics[width=\linewidth]{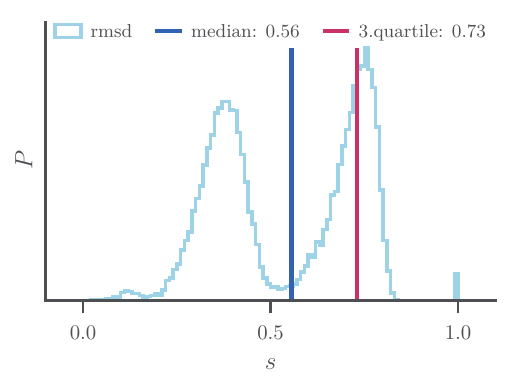}
		\caption{set A}
	\end{subfigure}
	\begin{subfigure}{0.3\linewidth}
		\includegraphics[width=\linewidth]{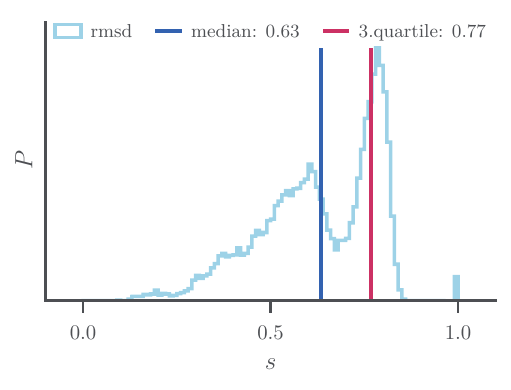}
	\caption{set B}
	\end{subfigure}
	\begin{subfigure}{0.3\linewidth}
		\includegraphics[width=\linewidth]{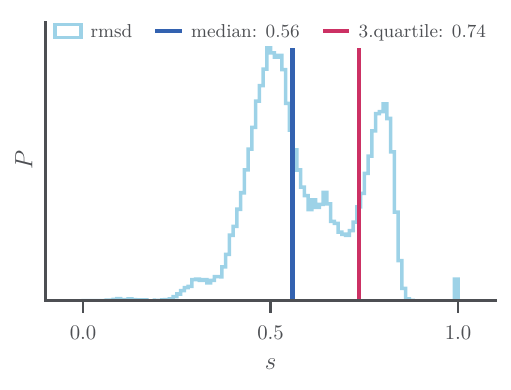}
	\caption{set C}
	\end{subfigure}
	\caption{Similarity distribution for RMSD-based similarity measure. The blue line indicates the median and the red line the third quartile of the distribution.}
    \label{fig:Stb_similaritydistribution}
\end{figure}

\subsubsection{conPCA}
\begin{figure}[H]
	\centering
	\begin{subfigure}{0.3\linewidth}
		\includegraphics[width=\linewidth]{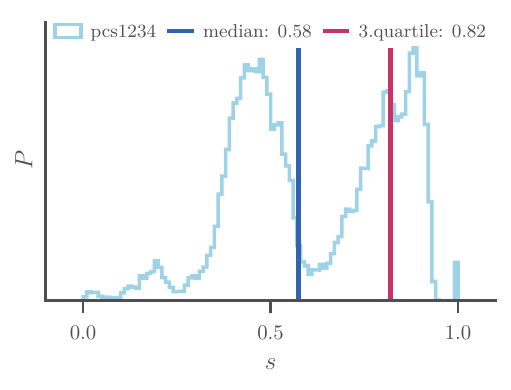}
		\caption{set A}
	\end{subfigure}
	\begin{subfigure}{0.3\linewidth}
		\includegraphics[width=\linewidth]{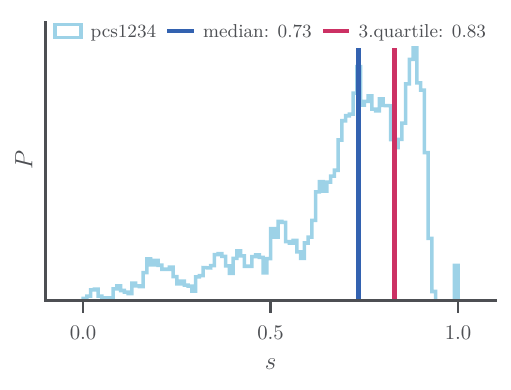}
		\caption{set B}
	\end{subfigure}
	\begin{subfigure}{0.3\linewidth}
		\includegraphics[width=\linewidth]{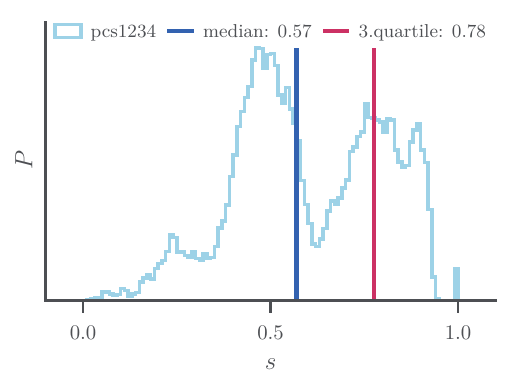}
		\caption{set C}
	\end{subfigure}
	\caption{Similarity distribution for contact PCA-based similarity measure. The blue line indicates the median and the red line the third quartile of the distribution.}
    \label{fig:Stb_conPCA1234_similarity_distribution}
\end{figure}

\subsection{Clustering results: Leiden/CPM and \textit{k}-medoids}
\subsubsection{RMSD}
\begin{figure}[H]
	\centering
	\begin{subfigure}{\linewidth}
	\includegraphics[width=\linewidth]{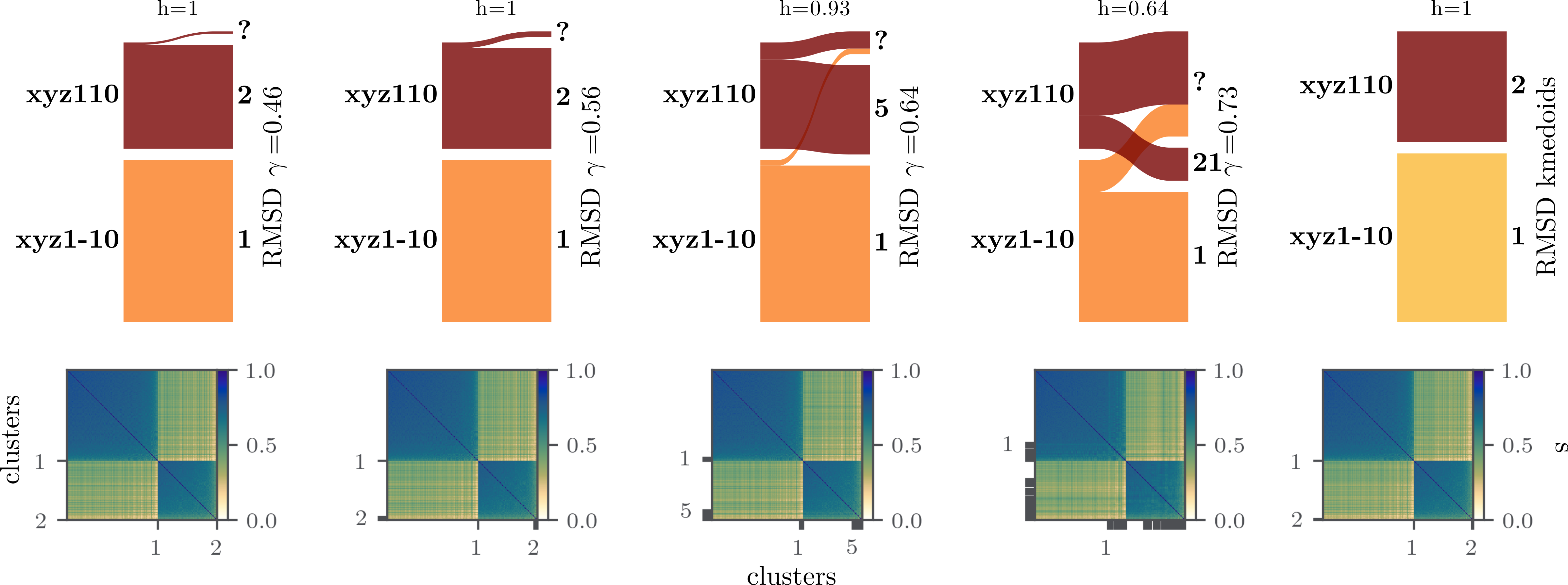}
	\caption{set A}
	\label{fig:Stb_setA_rmsd_sankeygrid}
    \end{subfigure}
    \begin{subfigure}{\linewidth}
	\includegraphics[width=\linewidth]{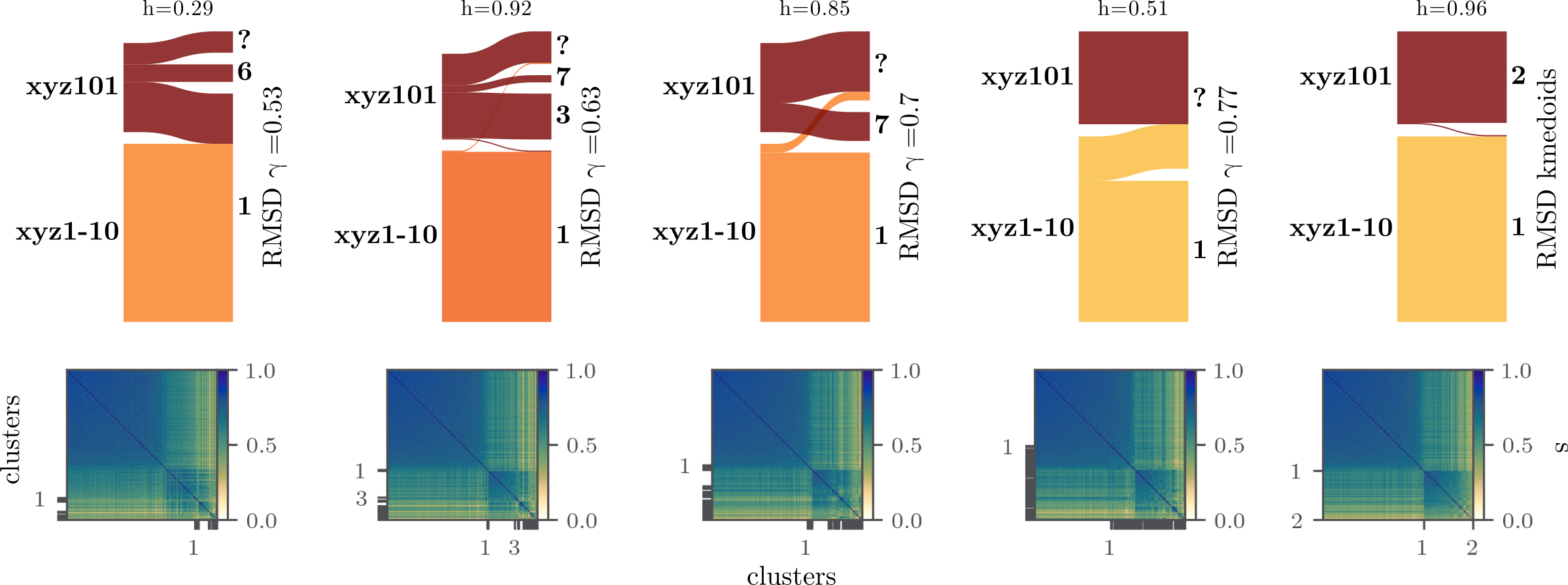}
	\caption{set B}
	\label{fig:Stb_setB_rmsd_sankeygrid}
    \end{subfigure}
    \begin{subfigure}{\linewidth}
	\includegraphics[width=\linewidth]{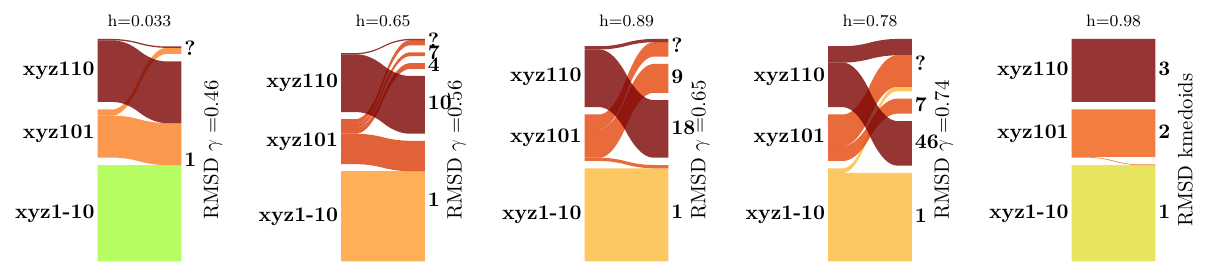}
	\caption{set C}
	\label{fig:Stb_setC_rmsd_sankeygrid}
 \end{subfigure}
    \caption{Sankey plots comparing RMSD clustering results for (a) top: set A, (b) top: set B  and (c) set C. On the left of each Sankey plot, the pulling vectors are shown. The $h$ on top shows the homogeneity score, the label on the right defines the input feature (RMSD) as well as the clustering method, $\gamma$ indicating Leiden/CPM clustering. (a) bottom and (b) bottom show the corresponding block similarity matrices.}
    \label{fig:Stb_rmsd_sankeygrid}
\end{figure}

\subsubsection{conPCA}
\begin{figure}[H]
	\centering
    \begin{subfigure}{\linewidth}
	\includegraphics[width=\linewidth]{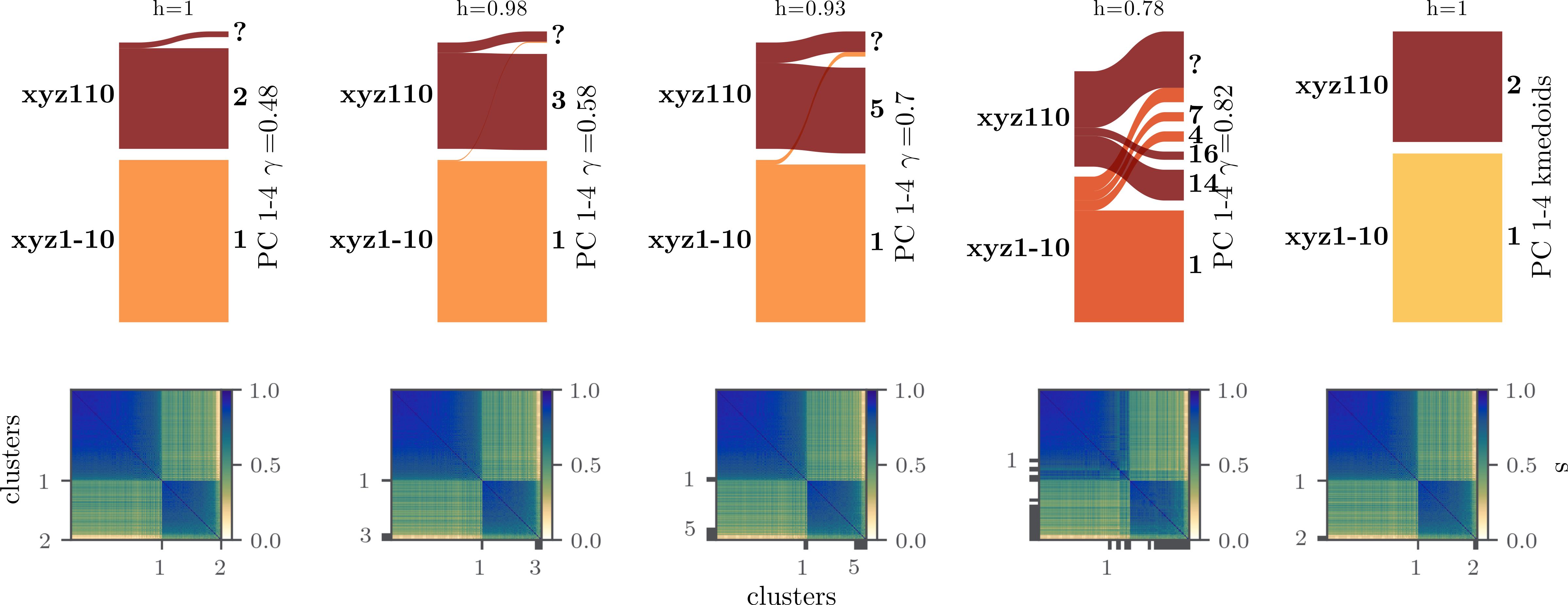}
	\caption{set A}
	\label{fig:Stb_setA_conPCA1234_sankeygrid}
    \end{subfigure}
    \begin{subfigure}{\linewidth}
	\includegraphics[width=\linewidth]{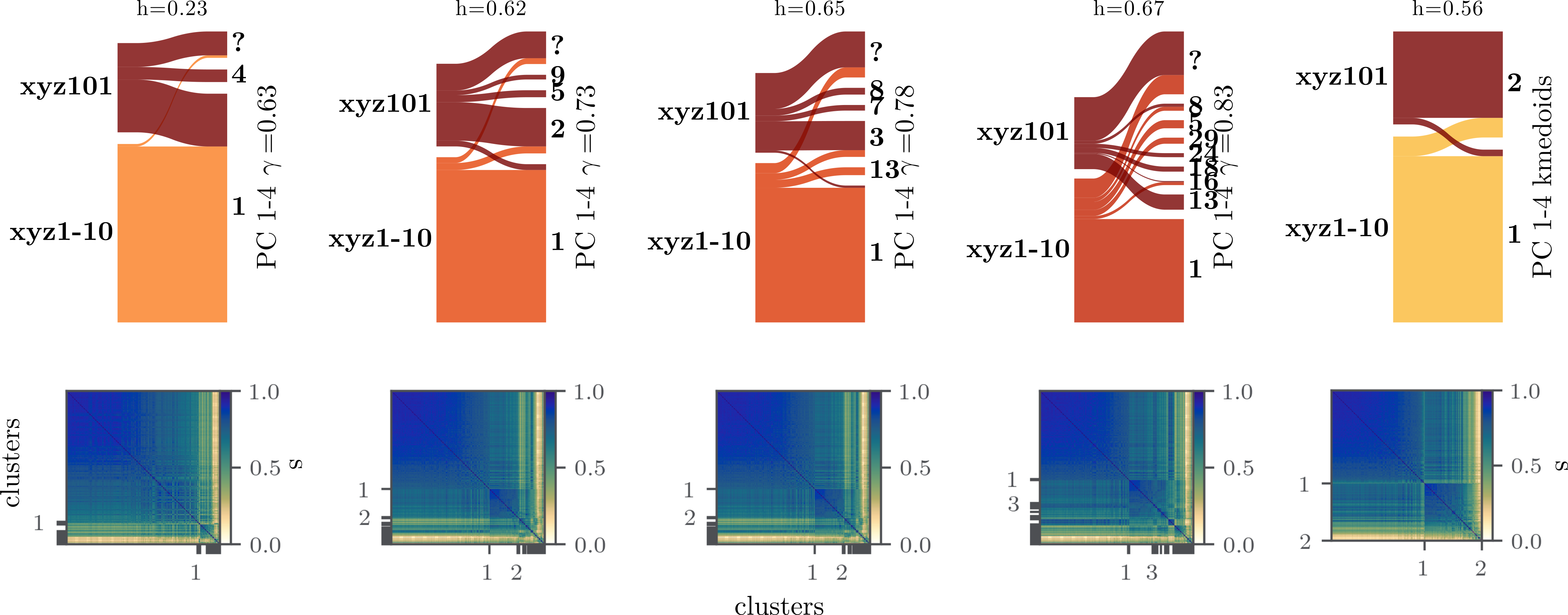}
	\caption{set B}
	\label{fig:Stb_setB_conPCA1234_sankeygrid}
    \end{subfigure}
    \begin{subfigure}{\linewidth}
	\includegraphics[width=\linewidth]{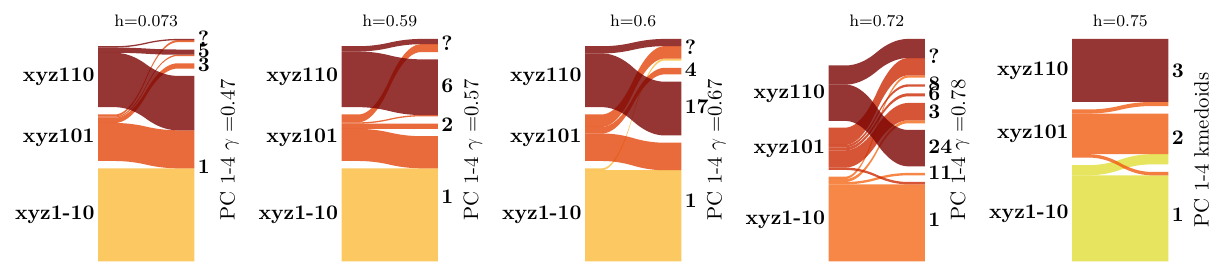}
	\caption{set C}
	\label{fig:Stb_setC_conPCA1234_sankeygrid}
    \end{subfigure}
    \caption{Sankey plots comparing conPCA clustering for (a) top: set A, (b) top: set B and (c) set C. On the left of each Sankey plot, the pulling vectors are shown. The $h$ on top shows the homogeneity score, the label on the right defines the input feature (conPCA 1-4) as well as the clustering method, $\gamma$ indicating Leiden/CPM clustering. (a) bottom and (b) bottom show the corresponding block similarity matrices.}
    \label{fig:Stb_conPCA1234_sankeygrid}
\end{figure}
\subsection{\textit{k}-medoids Silhouette scores}
\subsubsection{RMSD}
\begin{figure}[H]
	\centering
	\includegraphics[width=\linewidth]{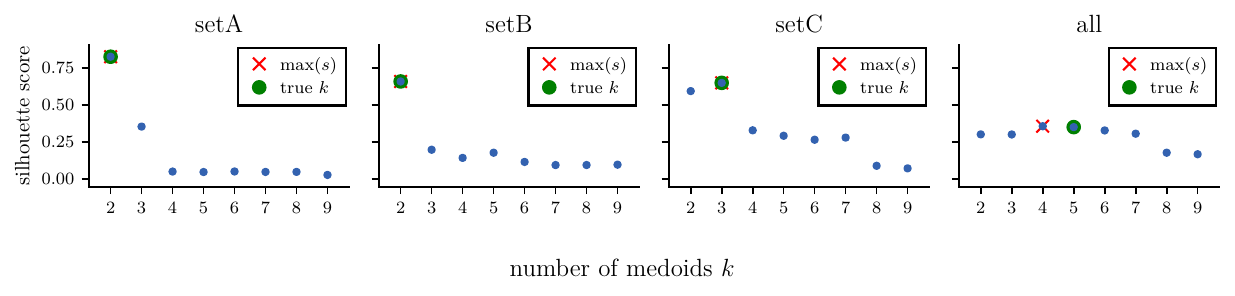}
	\caption{Silhouette scores of different St-b subsets clustered using a RMSD based similarity matrix with \textit{k}-medoids. The subset is indicated in the subplot title. The red cross marks the maximal Silhouette score, while the green dot marks the true number of clusters.}
	\label{fig:Stb_rmsd_kmedois_silouettescore}
\end{figure}

\subsubsection{conPCA}
\begin{figure}[H]
	\centering
	\includegraphics[width=\linewidth]{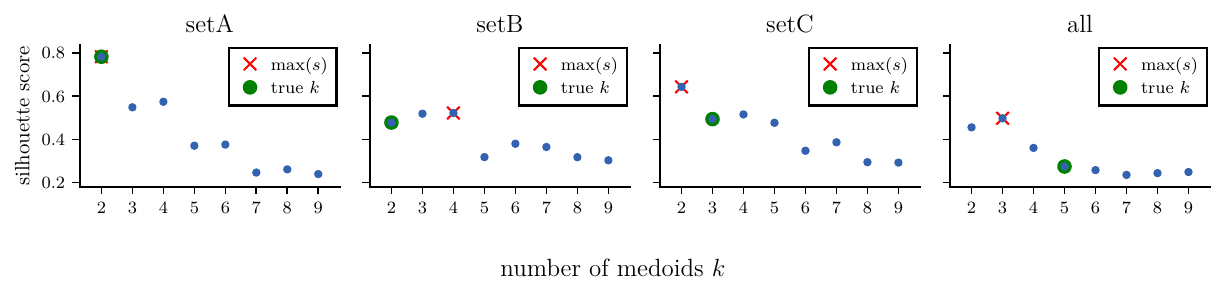}
	\caption{Silhouette scores of different St-b subsets clustered using a contact PC 1-4 based similarity matrix with \textit{k}-medoids. The subset is indicated in the subplot title. The red cross marks the maximal Silhouette score, while the green dot marks the true number of clusters.}
	\label{fig:Stb_conPCA1234_kmedois_silouettescore}
\end{figure}

\section{A\down{2a} adenosine receptor}
\subsection{Structural overview}
\begin{figure}[H]
	\centering
    \begin{subfigure}{0.45\linewidth}
        \includegraphics[width=\linewidth]{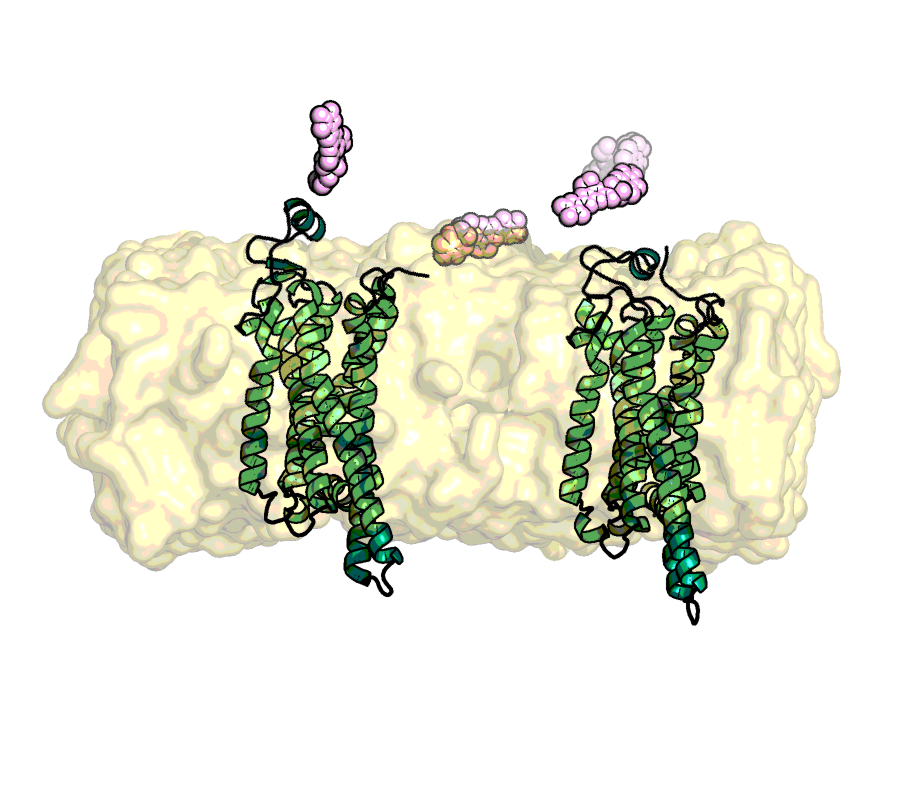}
        \caption{}
    \end{subfigure}
    \begin{subfigure}{0.3\linewidth}
        \includegraphics[width=\linewidth]{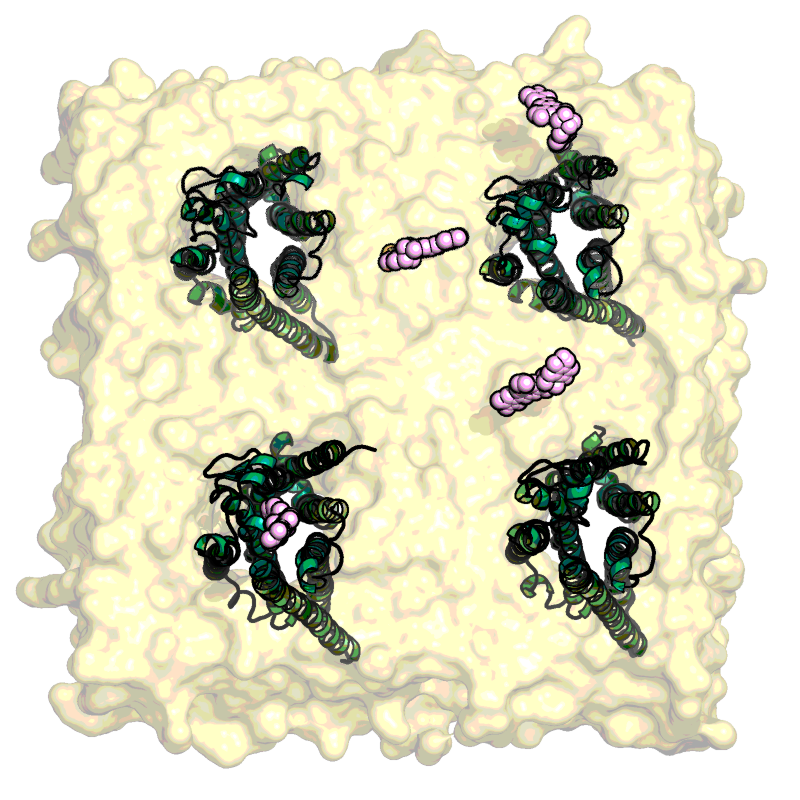}
        \caption{}
    \end{subfigure}
	%\label{fig:a2asystemrender}
    \begin{subfigure}{0.3\linewidth}
        \includegraphics[width=\linewidth]{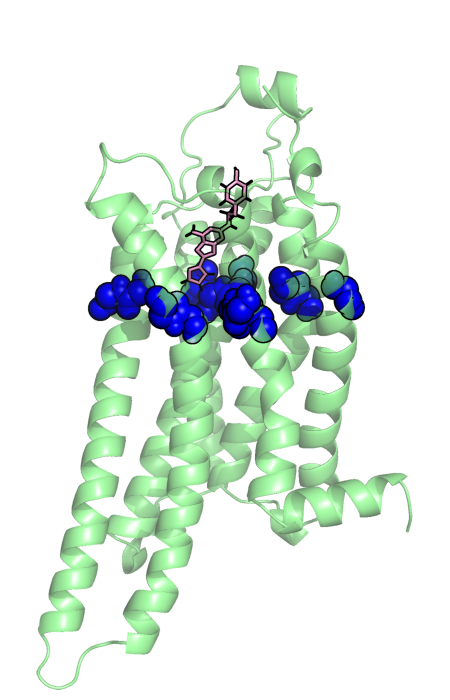}
        \caption{}
    \end{subfigure}
	%\label{fig:a2asystemrender}
    \begin{subfigure}{0.3\linewidth}
        \includegraphics[width=\linewidth]{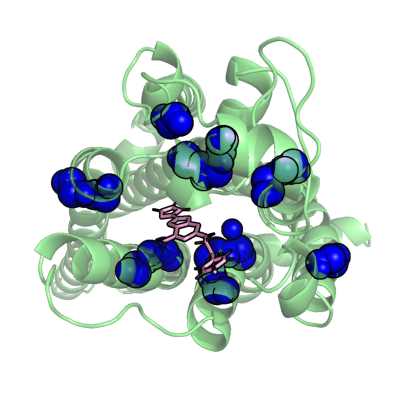}
        \caption{}
    \end{subfigure}
    \caption{A\down{2a} system. (a) and (b) visualization of simulation box with 4 proteins (green cartoon) with ligands (pink spheres) embedded in POPC bilayer (yellow volume) shown without water. (c)-(d) visualization of pull group residue backbone atoms (blue spheres): 14, 58, 59, 83, 84, 85, 136, 182, 183, 246, 247, 276, 277 and 278. Ligand as pink sticks.}
	\label{fig:a2asystemrender}
\end{figure}

\subsection{Importance of the PCs for conPCA similarities}
\begin{figure}[H]
	\centering
    \begin{subfigure}{0.7\linewidth}
        \includegraphics[width=\linewidth]{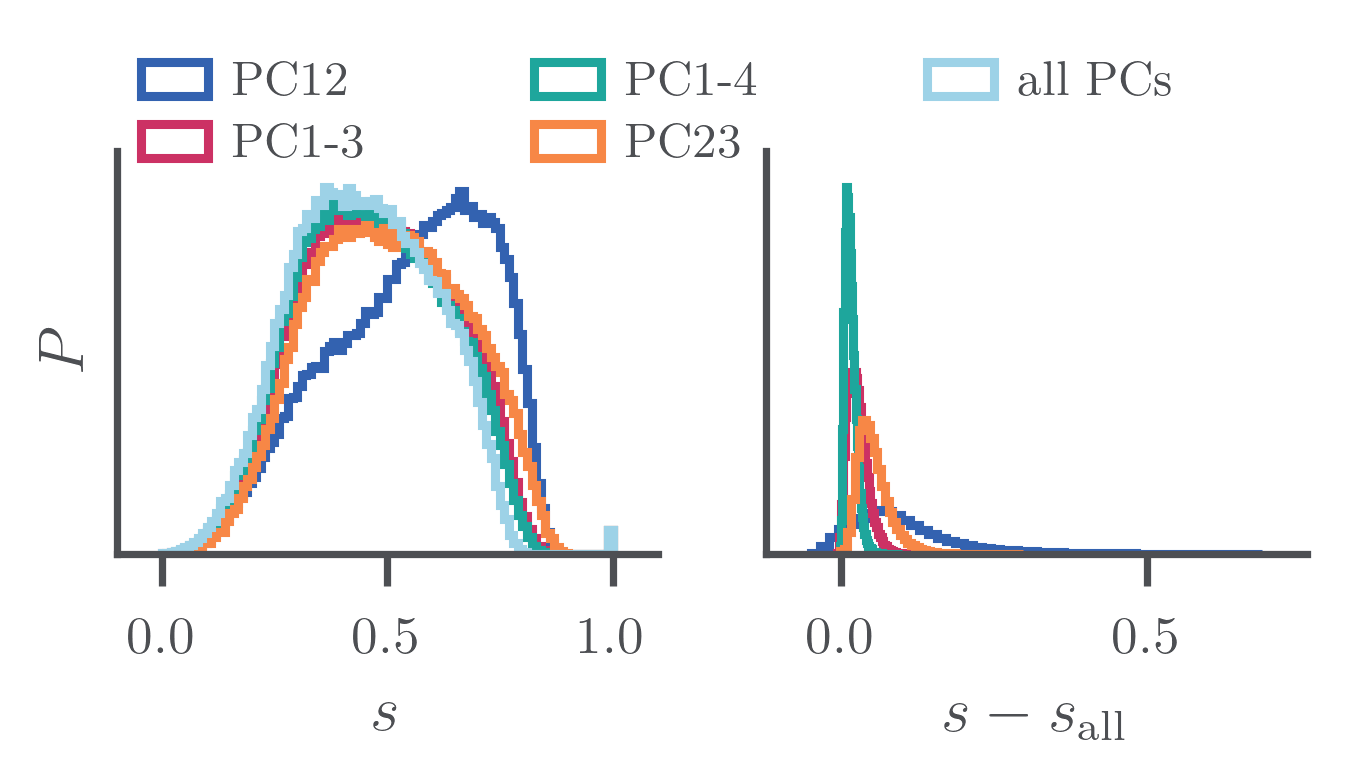}
    \end{subfigure}
	\caption{Left: probability distribution of the similarities $s$ calculated using different PCs. Right: probability distribution of the element-wise differences between similarities calculated based on PC subsets (e.g. PC12, PC1-3 etc.) and the full PCA space $s_\mathrm{all}$ (all PCs).}
	\label{fig:a2aconpcasimilaritydistribution}
\end{figure}
\subsection{Clustering results: Leiden/CPM}
\begin{figure}[H]
    \begin{subfigure}{0.45\textwidth}
        \includegraphics[width=\textwidth]{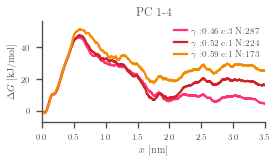}
        \caption{}
    \end{subfigure}
        \begin{subfigure}{0.45\textwidth}
        \includegraphics[width=\textwidth]{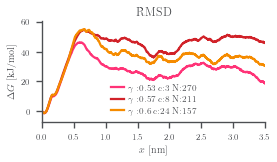}
        \caption{}
    \end{subfigure}
        \begin{subfigure}{0.3\linewidth}
        \includegraphics[width=\textwidth]{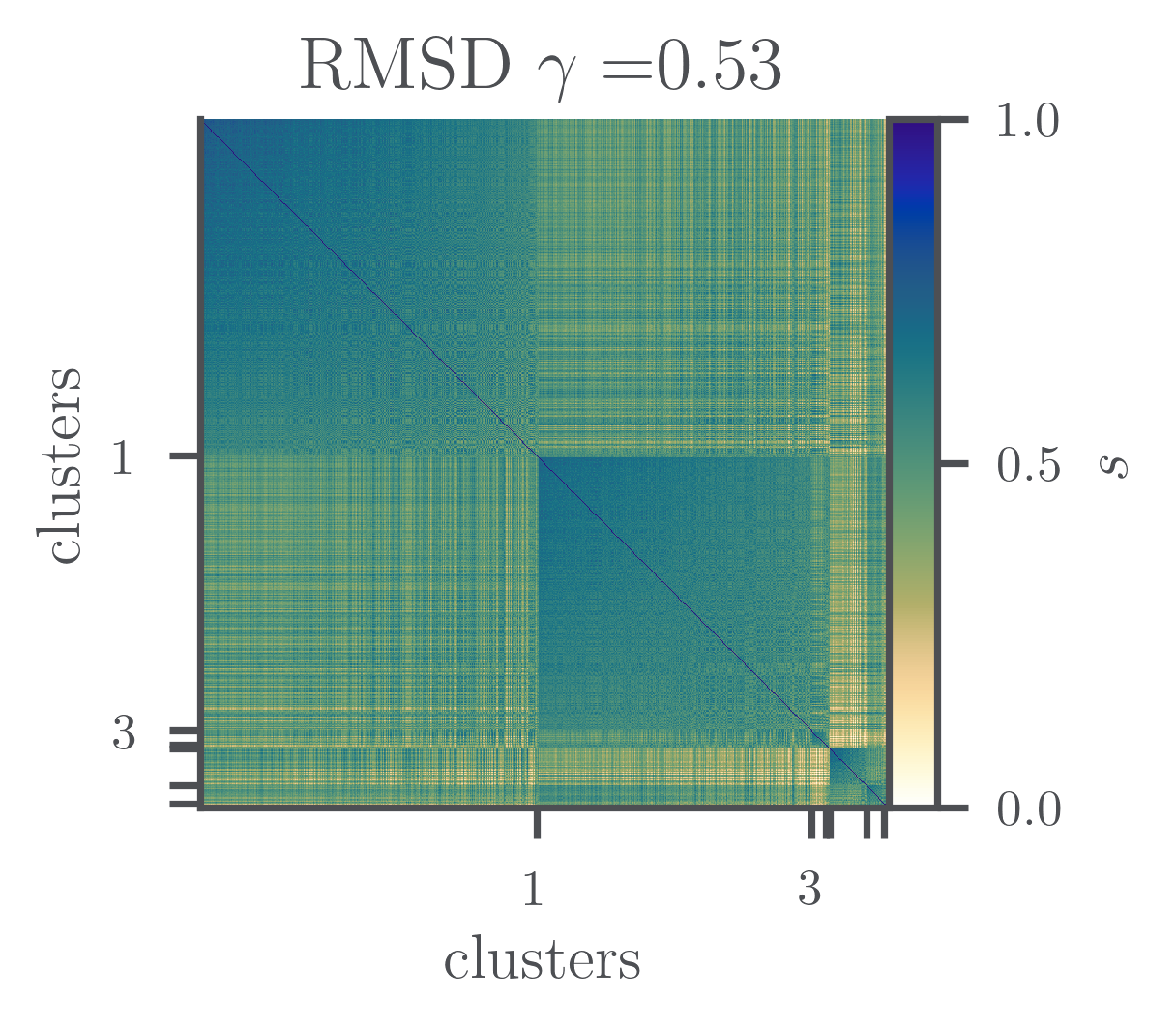}
        \caption{}
        %\label{fig:A2A_liglipdistmatrix}
    \end{subfigure}
    \caption{Estimated free energy of different (a) PC1-4 based and (b) RMSD-based Leiden/CPM clustering results. (c) RMSD-based similarity block matrix.}
    \label{fig:A2A_ZMA_leiden_dGprofile}
\end{figure}

%\subsubsection{Free energy estimation with bootstrapping errors}
\begin{figure}[H]
	\centering
	\includegraphics[width=\linewidth]{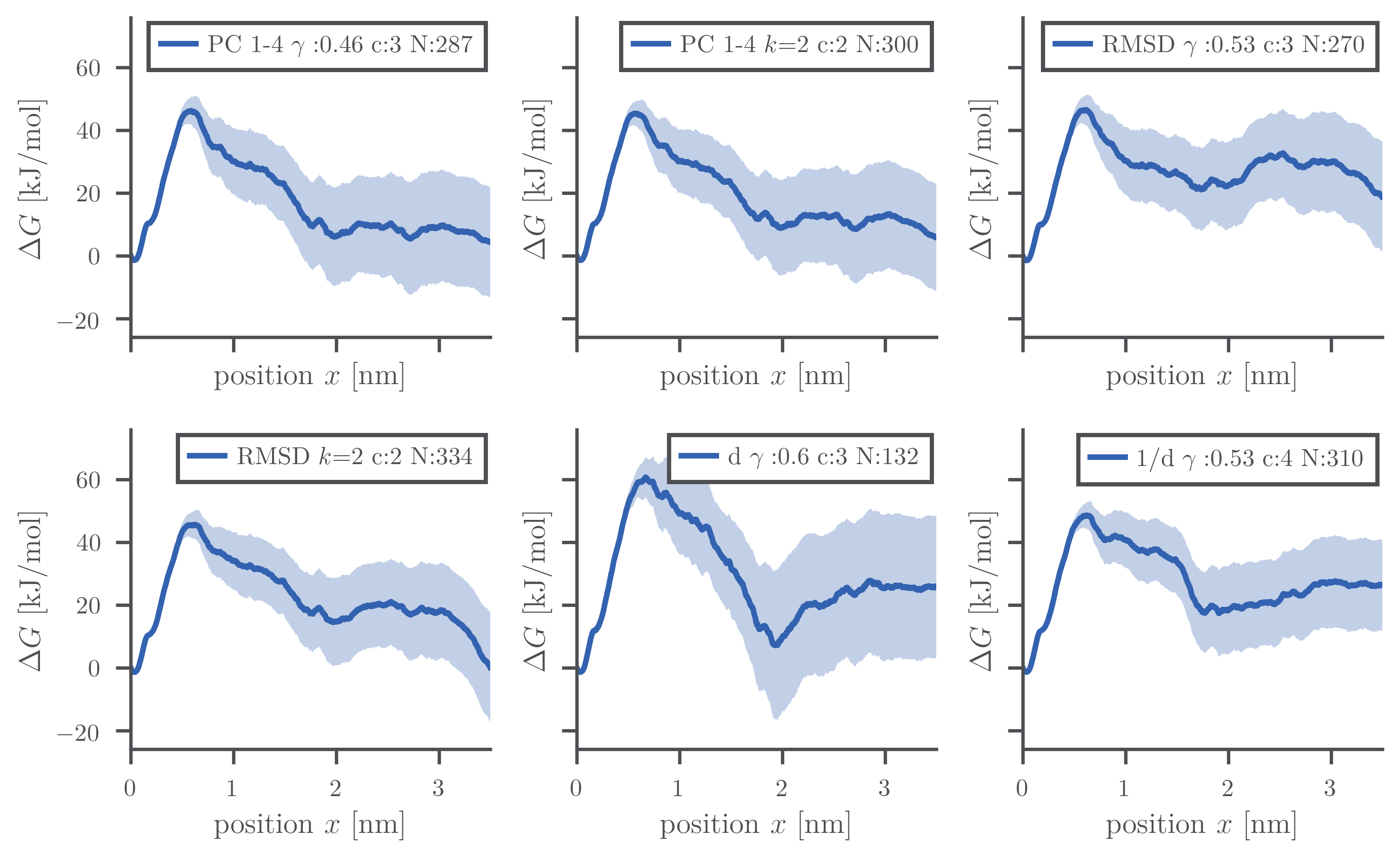}
	\caption{summary of the dcTMD results from the friction overestimation mitigating clusters. The label indicates the input feature (PC 1-4, RMSD, $d$ indicating $\delta_{\rm ligand-lipid}$ and $1/d$ indicating $1/\delta_{\rm ligand-lipid}$). In the legends, $\gamma$ denotes the used resolution parameters for Leiden/CPM based clustering and $k$ denotes the number of medoids for $k$-medoids clustering. $N$ indicates the number of trajectories in the cluster. The shaded area represents the standard deviation of 5000 bootstrapping resamples.}
	\label{fig:conpcarmsdvszma-popcdg}
\end{figure}
\begin{figure}[H]
    \centering
    \begin{subfigure}{0.9\textwidth}
    \includegraphics[width=\textwidth]{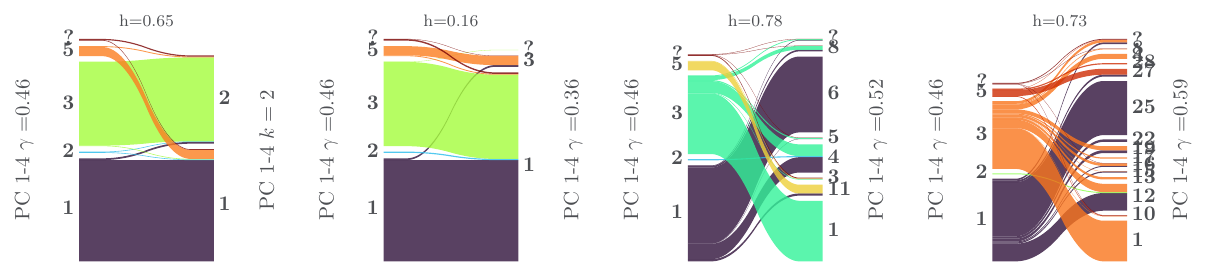}
    \end{subfigure}
    \begin{subfigure}{0.9\textwidth}
    \includegraphics[width=\textwidth]{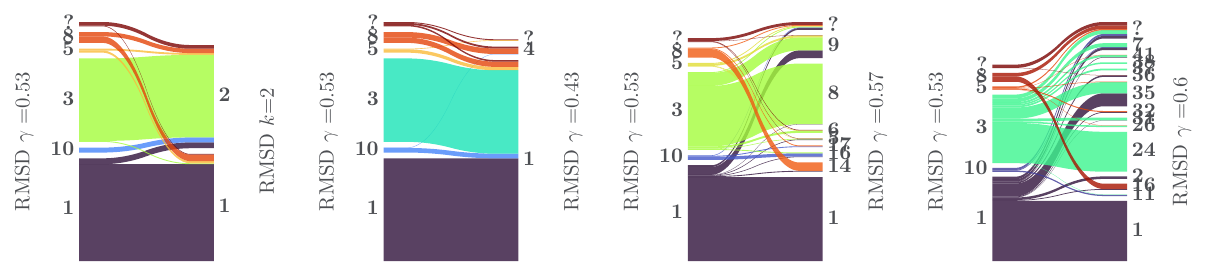}
    \end{subfigure}
    \caption{Sankey plots comparing contact PCs 1-4 based (top) and RMSD-based (bottom) Leiden/CPM clustering results using $\gamma \approx \text{median}(s_{ij})$ with results of \textit{k}-medoids and various $\gamma$.}
    \label{fig:A2A_ZMA_leiden_sankeygrid}
\end{figure}

\subsection{Clustering results: \textit{k}-medoids}

%The \textit{k}-medoids approach is tested next. Figure \ref{fig:kmedoids_block_dia} shows the block-diagonalized matrices (plots for higher \textit{k} are moved to \autoref{fig:appendix_kmedoids_block_dia}). Already for $k=4$ or even $k=3$, there are high inter-cluster similarities, a trend that increases with $k$. Having said that, the quality of the clusters themselves improves with increasing $k$, as expected. 

%The estimated free energies calculated for the different satisfactory clusters are shown in \autoref{fig:kmedoids_deltaG}. For $k=4$, the clusters 2 and 4 have been lumped and their result is shown. See \autoref{fig:appendix_kmedoids_deltaG} in the Appendix for the individual clusters and results for $k=5$. In all cases, the estimated free energies of the large (possibly lumped) clusters agree well with Leiden/CPM. Interestingly, the sizes of the clusters are very close to each other. This is explained by the sankey plots in Figs. \ref{fig:kmedoids_sankey_1} and \ref{fig:kmedoids_sankey_2} (more Sankeys plots are found in \autoref{fig:appendix_kmedoids_sankeys}). They confirm that on some occasions, clusters are simply split into two, as for example in the case of cluster 1 from $k=2$ if compared to $k=4$. In some other cases, clusters may even be preserved when increasing $k$.

\begin{figure}[H]
    \centering
    \begin{subfigure}[t]{.45\textwidth}
        \centering        
            \includegraphics[width=\linewidth]{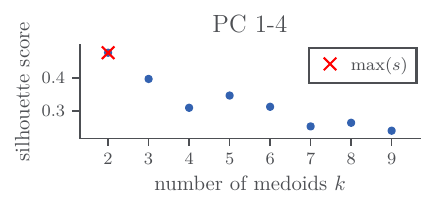}
        %\caption{}
        \label{fig:A2A_PC1234_kmedois_silouettescore}
    \end{subfigure}
    \begin{subfigure}[t]{.45\textwidth}
        \centering        
        \includegraphics[width=\linewidth]{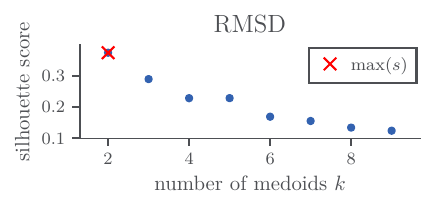}
        %\caption{}
        \label{fig:A2A_rmsd_kmedois_silouettescore}
    \end{subfigure}
    \caption{Silhouette scores of A\down{2a} $k$-medoids clustering results using a contact PCs 1-4 based
similarity matrix (left) and an RMSD-based similarity matrix (right). The red cross
marks the maximal Silhouette score.}
    \label{fig:A2A_kmedois_silouettescore}
\end{figure}

\begin{figure}[H]
    \centering
    \begin{subfigure}[t]{.75\textwidth}
        \centering
        \includegraphics[width=\linewidth]{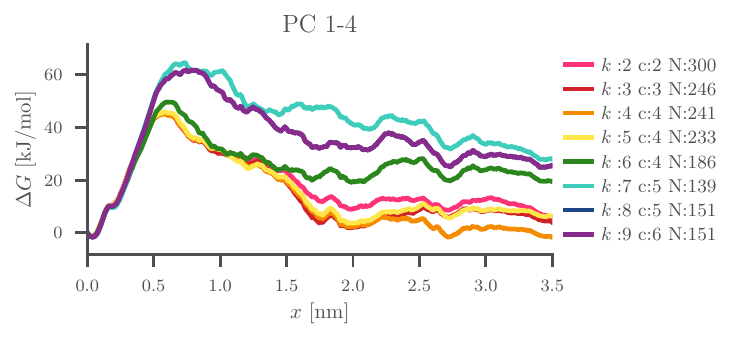}
        %\caption{}
        \label{fig:A2A_ZMA_SI_pc1234_dGprofile}
    \end{subfigure}
    \begin{subfigure}[t]{.75\textwidth}
        \centering
        \includegraphics[width=\linewidth]{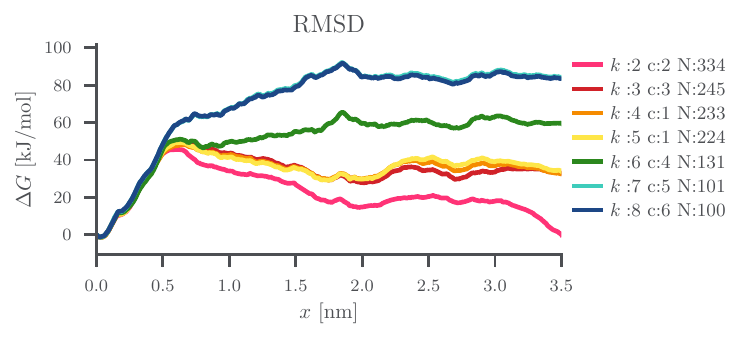}
        %\caption{}
        \label{fig:A2A_ZMA_SI_pc1234_kmedoids_dGprofile}
    \end{subfigure}
    \caption{Estimated free energies $\Delta G$ for A\down{2a} $k$-medoids clusters, contact PC 1-4 based (top) and RMSD based (bottom).}
    \label{fig:A2A_ZMA_kmedoids_dGprofile}
\end{figure}

\begin{figure}[H]
    \centering
    \includegraphics[width=.8\linewidth]{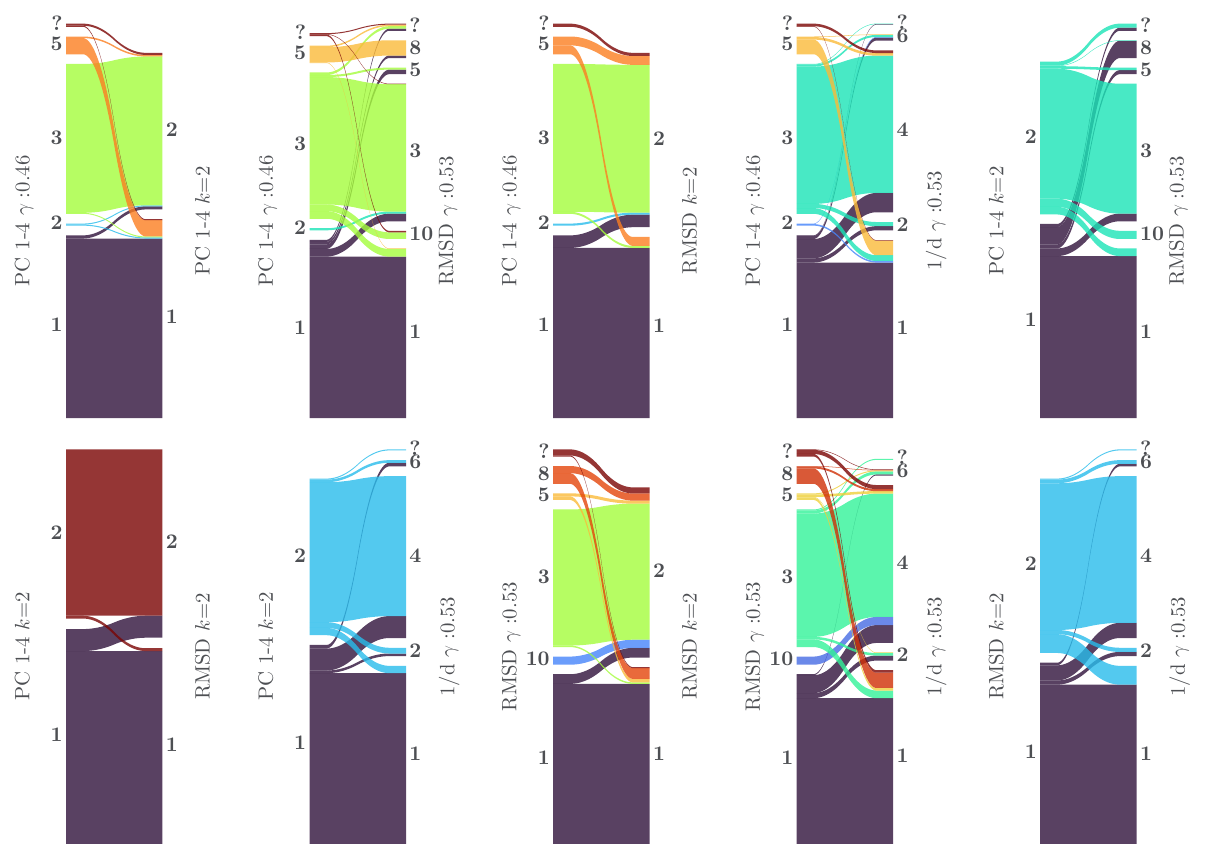}
    \caption{Sankey plots comparing CPM and \textit{k}-medoids Leiden clustering results using different input features: PC 1-4, RMSD, $d$ indicating $\delta_{\rm ligand-lipid}$ and $1/d$ indicating $1/\delta_{\rm ligand-lipid}$.}
    \label{fig:A2A_sankey}
\end{figure}

\subsection{Benchmarking of additional clustering methods in A\down{2a}}\label{chap:appendix_a2a_additional}

The following analyses were conducted on a reduced dataset with 344 trajectories. The comparisons to Leiden/CPM clusters are thus for illustrative purposes only. These clusters are not the same as the ones in the main text, but hold the same interpretation, just for the smaller dataset.

\subsubsection{Leiden/Modularity}

The Constant Potts Model is one possible objective function for the Leiden community detection algorithm. An alternative is Modularity, which is defined as\cite{Diez2022} 
\begin{equation}
    \phi_\text{mod} = \frac{1}{2m} \sum_c \left( e_c - \frac{k_c^2}{2m} \right)
\end{equation}
where we sum over all clusters $c$, the sum of similarities in a cluster is $e_c$, $k_c$ counts the similarities, i.e., graph edges, in cluster $c$ and $m$ denotes the total number of edges in the graph. Notably, this formulation of Modularity does not feature any tunable parameter that would play a similar role as $\gamma$ for the Constant Potts Model. 

\begin{figure}[H]
    \centering
    \begin{subfigure}[t]{.49\textwidth}
        \centering        
        \includegraphics[width=\linewidth]{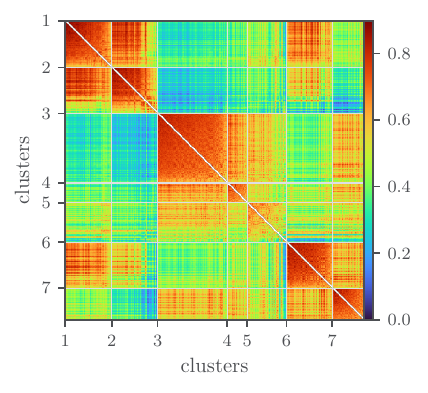}
        \caption{}
        \label{fig:mod_sim_mat}
    \end{subfigure}
    \begin{subfigure}[t]{.49\textwidth}
        \centering
        \includegraphics[width=.9\linewidth]{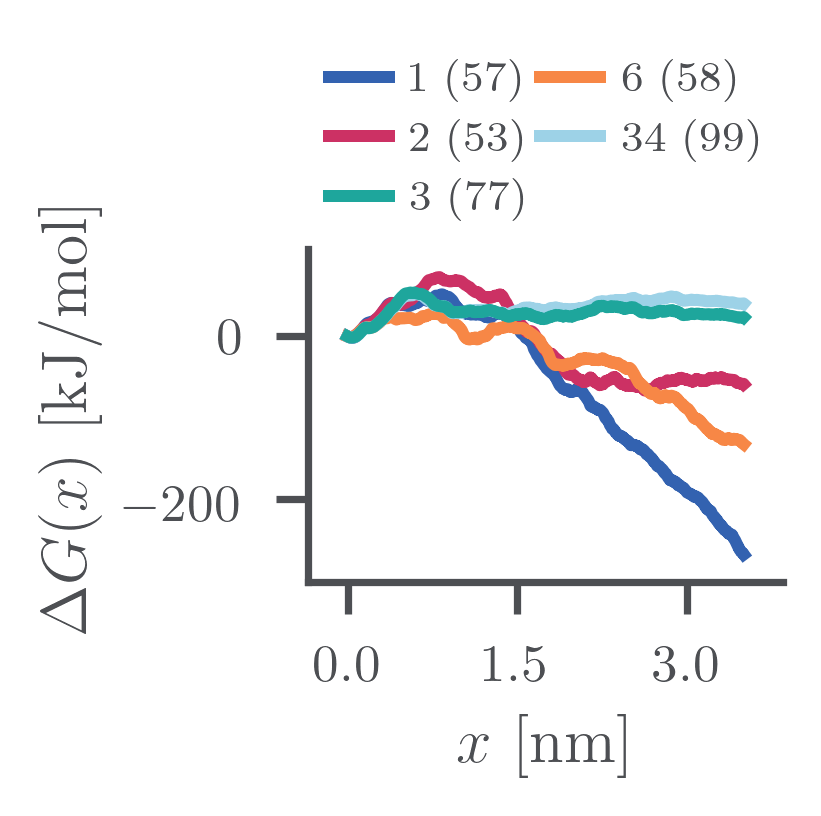}
        \caption{}
        \label{fig:mod_dG}
    \end{subfigure}
    \caption{Results from Leiden/modularity. (a) Similarity block matrix. (b) Estimated free energies for the clusters with size $N>50$. The cluster $34$ is lumped from 3 and 4. The number of trajectories in a cluster is in parentheses.}
\end{figure}

The similarity matrix in Fig.~\ref{fig:mod_sim_mat} reveals seven clusters, some of which have high inter-cluster similarities. Especially clusters 1 and 2 lead to the suspicion that there should be fewer clusters than suggested by the Modularity objective function. The estimated free energies of all clusters of size $N>50$ are shown in Fig.~\ref{fig:mod_dG}. Cluster 2 seems promising and can be lumped with cluster 3 without any major impact. However, even the lumped cluster is comparatively small. Other amalgamations do not yield improved results with regard to friction overestimation. Overall, our results agree with the literature\cite{Diez2022,Traag2019} that Modularity performs worse than the Constant Potts Model.

\subsubsection{Complete linkage}

\begin{figure}[H]
    \centering
    \begin{subfigure}[t]{.95\textwidth}
        \centering
            \includegraphics[width=.45\linewidth]{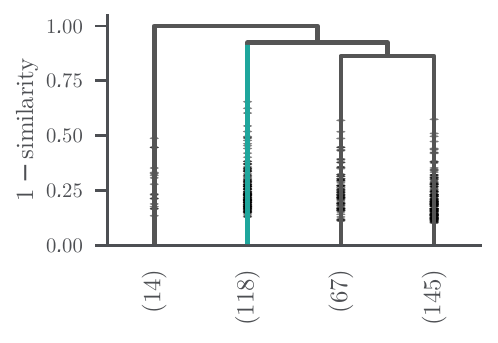}
        \caption{}
        \label{fig:clinkage_dendro}
    \end{subfigure}
    \begin{subfigure}[t]{.32\textwidth}
        \centering
        \includegraphics[width=\linewidth]{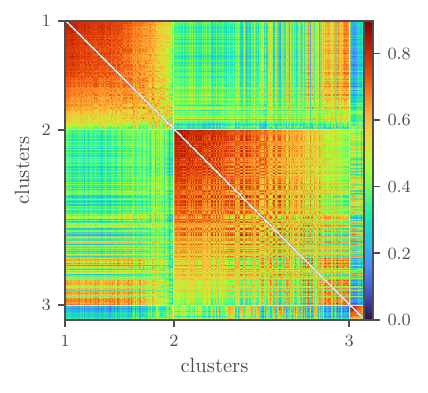}
        \caption{$\gamma_\mathrm{cl} = 0.92$}
    \end{subfigure}
    \hfill
    \begin{subfigure}[t]{.32\textwidth}
        \centering
        \includegraphics[width=\linewidth]{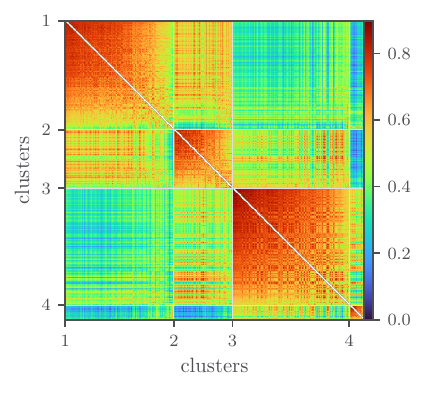}
        \caption{$\gamma_\mathrm{cl} = 0.85$}
        \label{fig:CL_block_dia_b}
    \end{subfigure}
    \hfill
    \begin{subfigure}[t]{.32\textwidth}
        \centering
        \includegraphics[width=\linewidth]{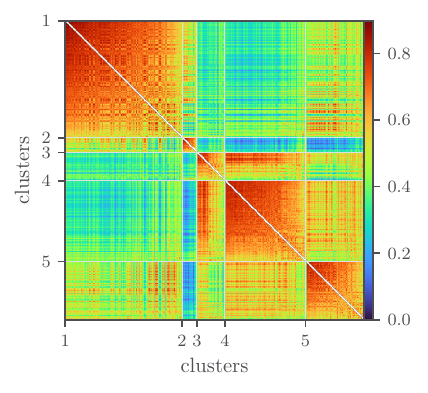}
        \caption{$\gamma_\mathrm{cl} = 0.70$}
    \end{subfigure}
    \caption{Results from complete linkage. (a) truncated dendrogram. (b)-(d) similarity block matrices for different resolution parameters $\gamma_\mathrm{cl}$.}
    \label{fig:CL_block_dia}
\end{figure}

Next, the hierarchical method complete linkage is evaluated. A truncated dendrogram is shown in Fig.~\ref{fig:clinkage_dendro}. Figure~\ref{fig:CL_block_dia}\,b-d show the similarity block matrices for different values of the cutoff parameter $\gamma_\mathrm{cl}$ around the similarities where the splits in the dendrogram appear. Lower values quickly lead to many clusters with high inter-cluster similarity, which are thus not shown.

\begin{figure}[H]
    \centering
    \begin{subfigure}[t]{.40\textwidth}
        \centering        
        \includegraphics[width=\linewidth]{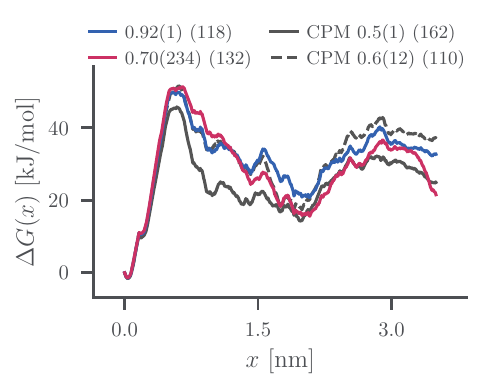}
        \caption{}
        \label{fig:cl_dG}
    \end{subfigure}
    \hfill
    \begin{subfigure}[t]{.29\textwidth}
        \centering
        \includegraphics[width=.8\linewidth]{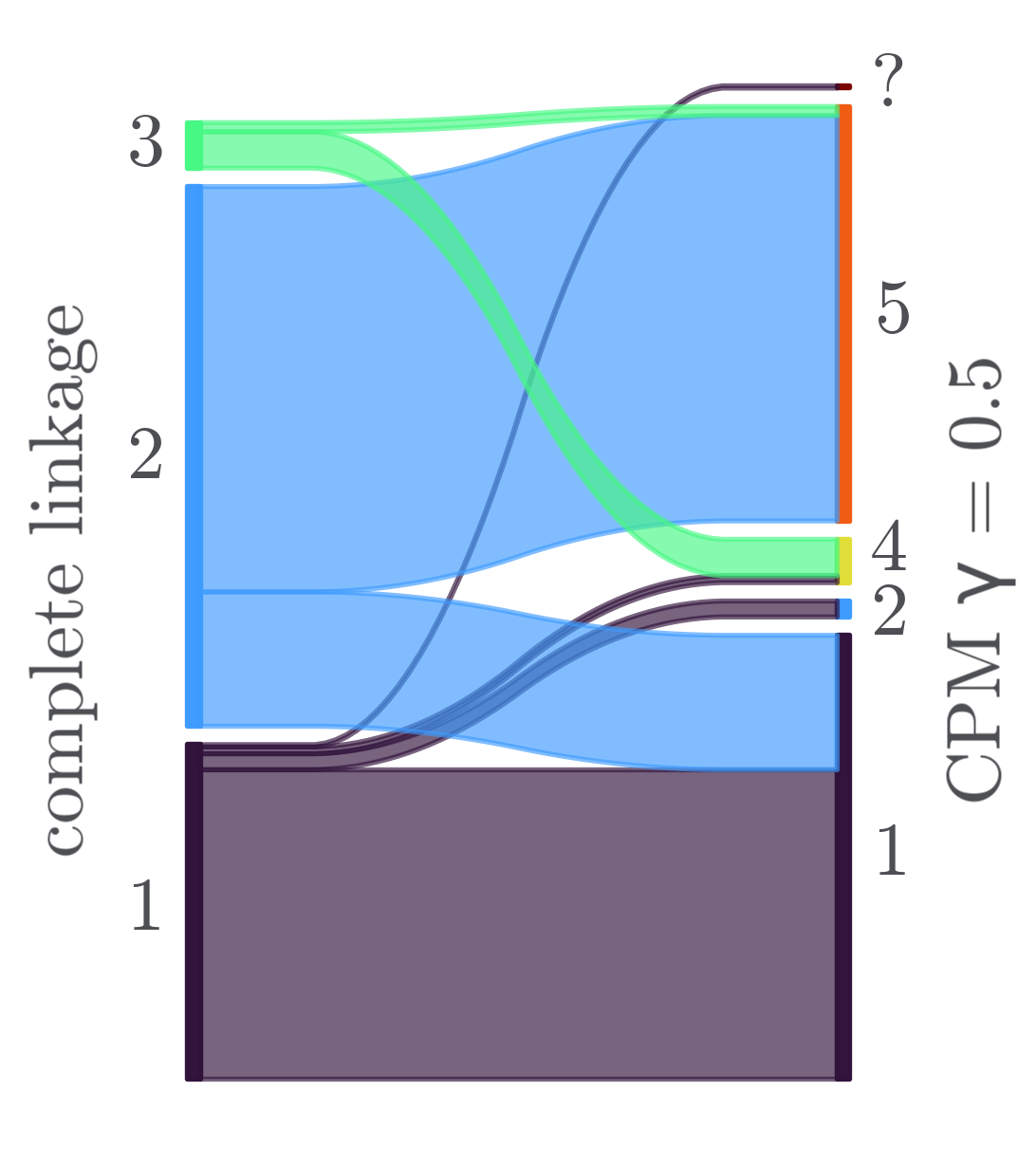}
        \caption{}
        \label{fig:cl_CPM_0.5_sankey}
    \end{subfigure}
    \hfill
    \begin{subfigure}[t]{.29\textwidth}
        \centering
        \includegraphics[width=.8\linewidth]{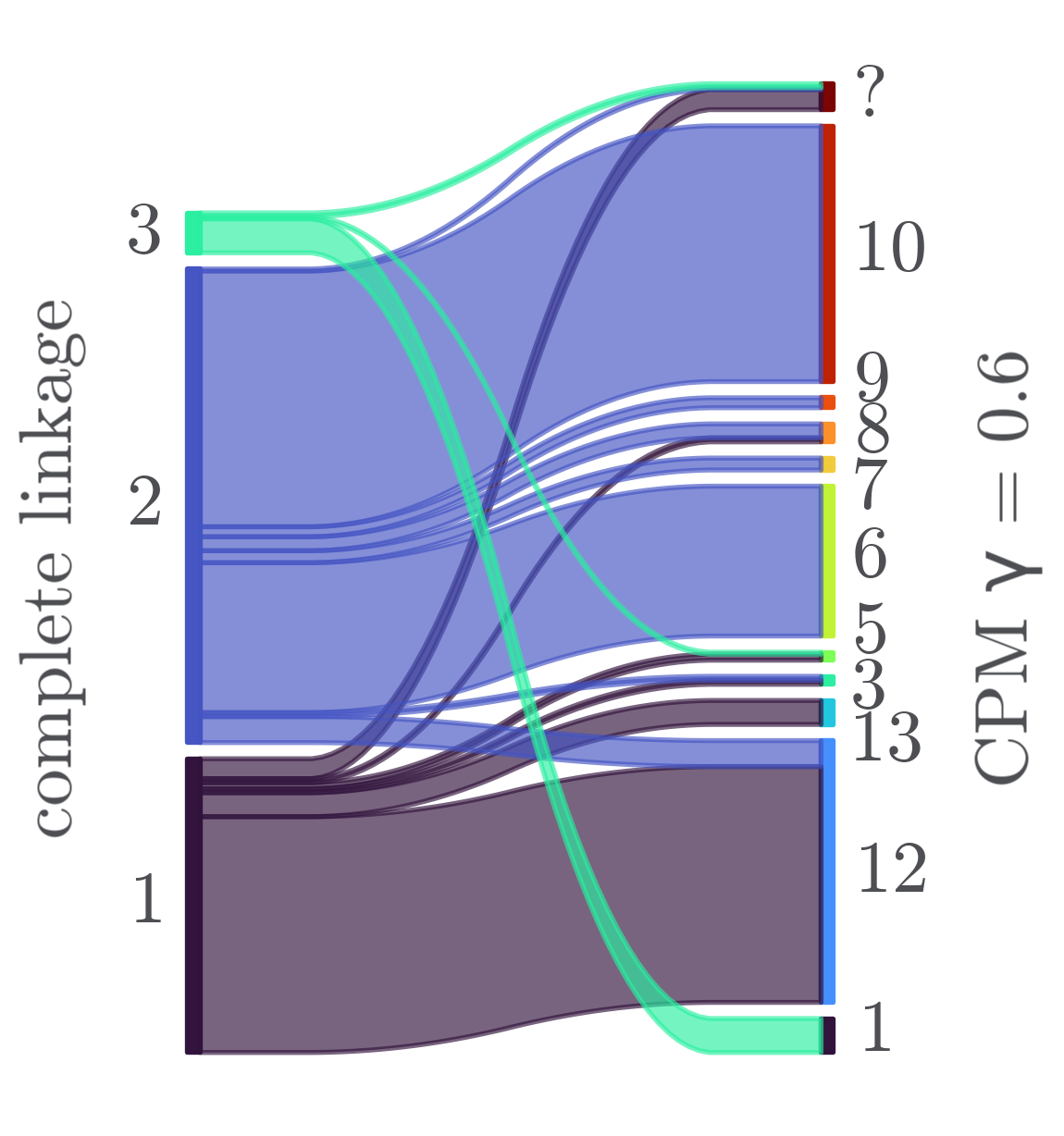}
        \caption{}
        \label{fig:cl_CPM_0.6_sankey}
    \end{subfigure}
    \caption{(a) Comparison of the estimated free energy $\Delta G$ of satisfactory clusters from complete linkage and Leiden/CPM with $\gamma=0.5, 0.6$. Legend: $\gamma_\text{cl}$/$\gamma$ (cluster number(s)) (population). (b) Sankey plot between the clusters from complete linkage with $\gamma_\mathrm{cl} = 0.92$ and Leiden/CPM with $\gamma=0.5$. (c) Sankey plot between the clusters from complete linkage with $\gamma_\mathrm{cl} = 0.92$ and Leiden/CPM with $\gamma=0.6$. Quotation marks (``?'') bundle small clusters.}
\end{figure}

A cluster without friction overestimation artefact, in the following denoted as ''satisfactory'' cluster, first appears for $\gamma_\mathrm{cl} = 0.92$. It is marked in green in Fig.~\ref{fig:clinkage_dendro}. This cluster 1 can be found again for $\gamma_\mathrm{cl} = 0.85$ in Fig.~\ref{fig:CL_block_dia_b} as cluster 3. For $\gamma_\mathrm{cl} = 0.70$, it is split into the clusters 3 and 4. The smaller cluster 2 can be added, which still yields a reasonable estimated free energy, as Fig.~\ref{fig:cl_dG} shows. There, the satisfactory clusters from Leiden/CPM with $\gamma=0.5 \text{ and } 0.6$ serve as comparison. While the curves are quite similar, one of the CPM clusters is notably larger. However, with complete linkage, the shown results are the largest clusters that are reasonably possible. The Sankey plots in Figs.~\ref{fig:cl_CPM_0.5_sankey} and \ref{fig:cl_CPM_0.6_sankey} show that the satisfactory linkage cluster $1$ is mostly preserved as cluster $12$ for Leiden/CPM with $\gamma=0.6$.

% \subsubsection{Comparison of RMSD and and conPCA-based trajectory clusters}

% \begin{figure}[H]
%     \centering
% \begin{subfigure}{0.45\textwidth}
%                 \includegraphics[width=\textwidth]{a2a/compare_conPCA23_RMSD_edited}
%         \caption{}
%     \end{subfigure}
% \begin{subfigure}{0.45\textwidth}
%                 \includegraphics[width=\textwidth]{a2a/compare_conPCA23_RMSDNN_edited}
%         \caption{}
%     \end{subfigure}
% \begin{subfigure}{0.45\textwidth}
%                 \includegraphics[width=\textwidth]{a2a/compare_RMSD_RMSDNN_edited}
%         \caption{}
%     \end{subfigure}
%     \caption{Sankey plots comparing RMSD and conPCA clusters. Quotation marks (``?'') bundle small clusters (Leiden) or unassigned trajectories (NeighborNet).}
%     \label{fig:a2a_sankey_RMSD_Leiden}
% \end{figure}

\subsubsection{NeighborNet}
The full dendrograms of linkage clustering can be difficult to assess visually.  This issue is improved for the NeighborNet\cite{Bryant2002} networks, which present the entire dissimilarity matrix, and 
additionally take into account ambiguity in the data. This inspired a prior study from our group\cite{Bray2022}, in which we recommended NeighborNet for trajectory clustering based on RMSD dissimilarities over dendrograms. 

\begin{figure}[H]
    \centering
    \includegraphics[width=.65\textwidth]{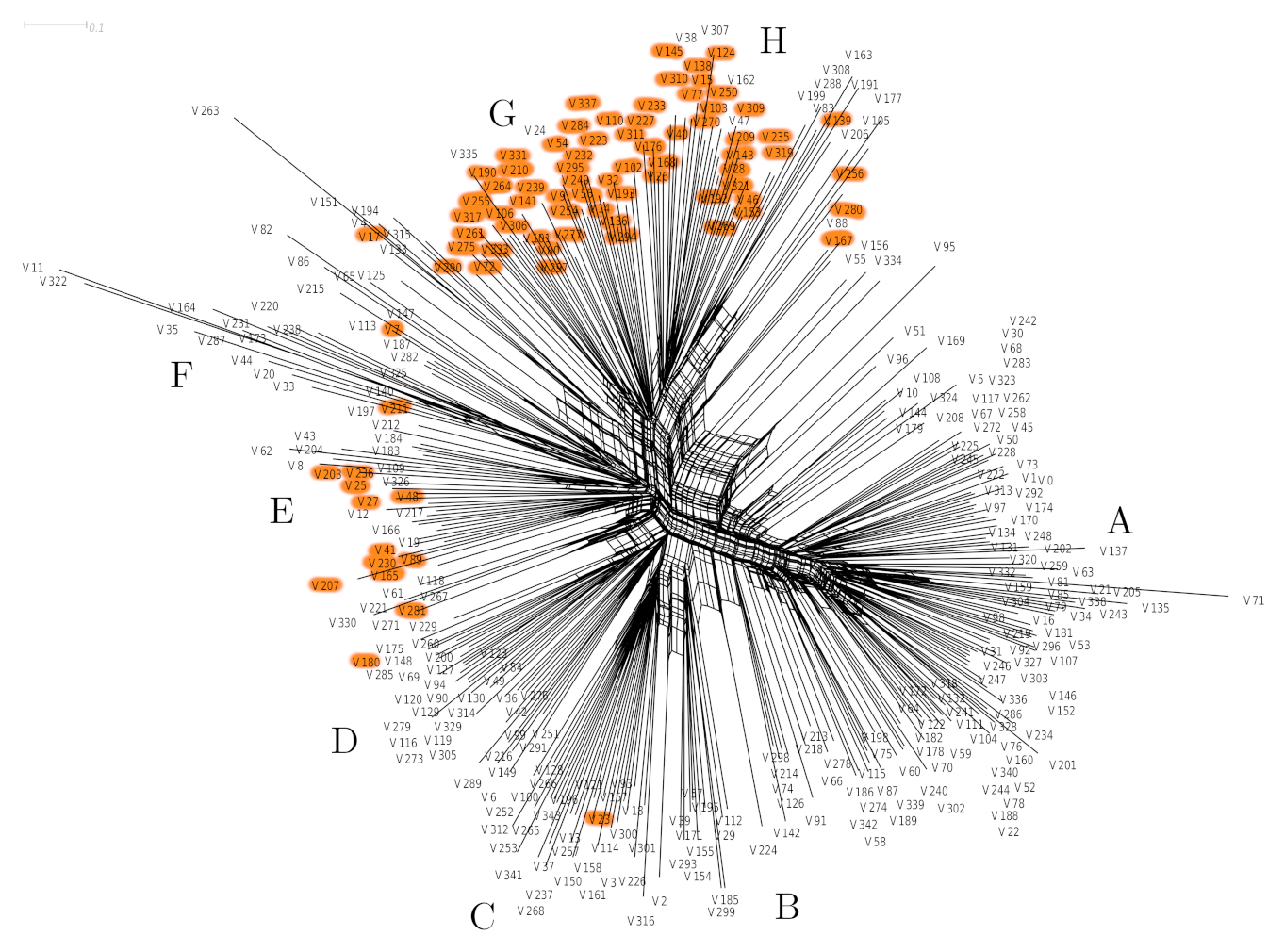}
    \caption{Comparing the RMSD NeighborNet and an RMSD Leiden/CPM for a $\gamma$ similar to the corresponding one from the main text. The trajectories from the Leiden clusters are marked in orange on the NeighborNet. The figure was generated with SplitsTree\cite{Huson2006}.}
    \label{fig:a2a_leiden_on_NN}
\end{figure}

One downside of NeighborNet, however, is the need for tedious user input and comparably complex decisions, which forbids scaling the workflow. What weights even more is a second disadvantage, which we find in this study and that is visualized in Fig.~\ref{fig:a2a_leiden_on_NN}. The NeighborNet splits graph from RMSD dissimilarities in the A\down{2a} complex is presented, and the trajectories corresponding to a cluster from Leiden clustering similar to the ones from this study are highlighted in orange. Two points stand out: Firstly, not all nodes in the vicinity of the regions G and H are orange. Notably, the outermost trajectories denote large dissimilarities, which is why Leiden/CPM decides to exclude them from the cluster.  Secondly, some isolated trajectories in the vicinity of C, D, E and F belong to the Leiden cluster as well. They often lie close to the cobweb-like middle of the network. This means that their distance to the cluster G or H is rather small, which is why they are more similar to G and H than the aforementioned outermost nodes. In other words, for a node, another node on the other side of the network might be closer than what appear to be its neighbors. For the pathway separation workflow, this behavior of NeighborNet is clearly undesirable.

\newpage

\subsection{A\down{2a} ligand characteristics}
\subsubsection{Trajectory sorting via ligand lipid distances}
\paragraph{Similarity calculation:}
Euclidean distances are used to compare the ligand-lipid distance $\delta^\text{lig-lip}_i$ and inverse distance time traces $1/\delta^\text{lig-lip}_i$, respectively, of different ligand unbinding runs via
\begin{align*}
	d^\text{lig-lip}_{ij} &= \left<\sqrt{\left(\delta^\text{lig-lip}_i(t) - \delta^\text{lig-lip}_j(t)\right)^2}\right>_t,\\
	d^\text{inv lig-lip}_{ij} &= \left<\sqrt{\left(1/\delta^\text{lig-lip}_i(t) - 1/\delta^\text{lig-lip}_j(t)\right)^2}\right>_t.
 \end{align*}
To reduce it to a scalar, time-independent quantity, we average over the entire simulation time.
\begin{figure}[H]
    \centering
    \begin{subfigure}{0.45\linewidth}
        \includegraphics[width=\textwidth]{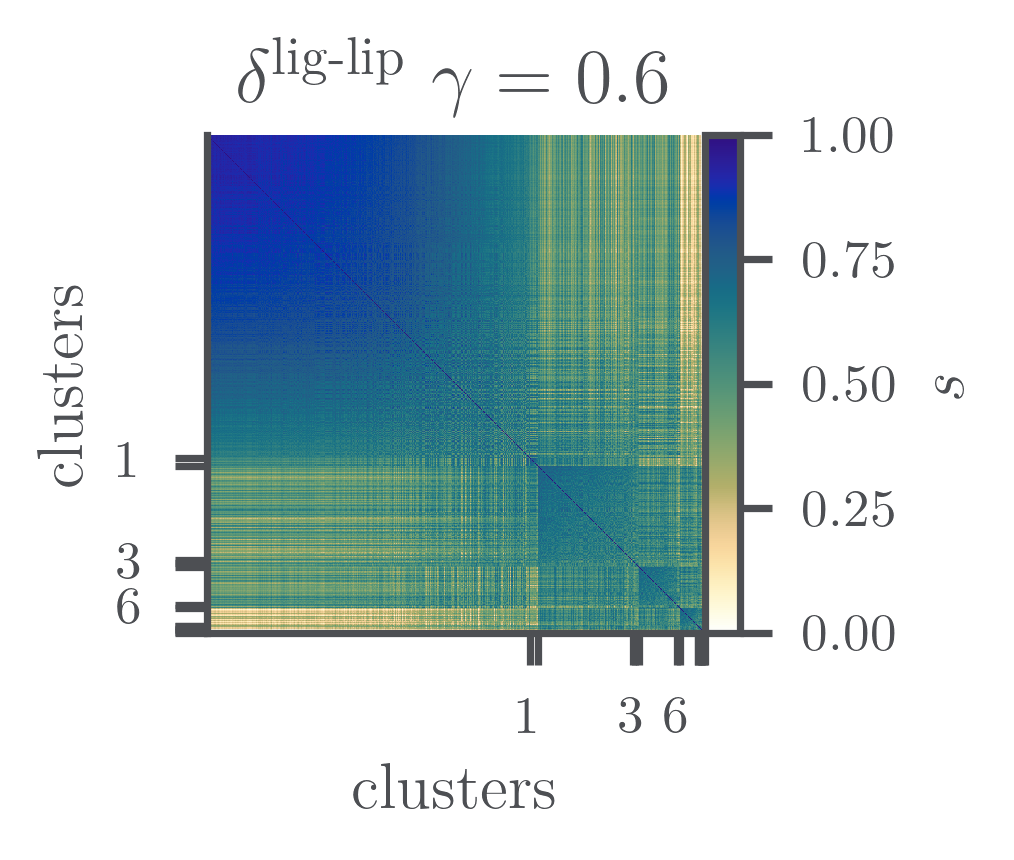}
        \caption{}
        \label{fig:A2A_liglipdistmatrix_dist}
    \end{subfigure}
    \begin{subfigure}{0.45\linewidth}
        \includegraphics[width=\textwidth]{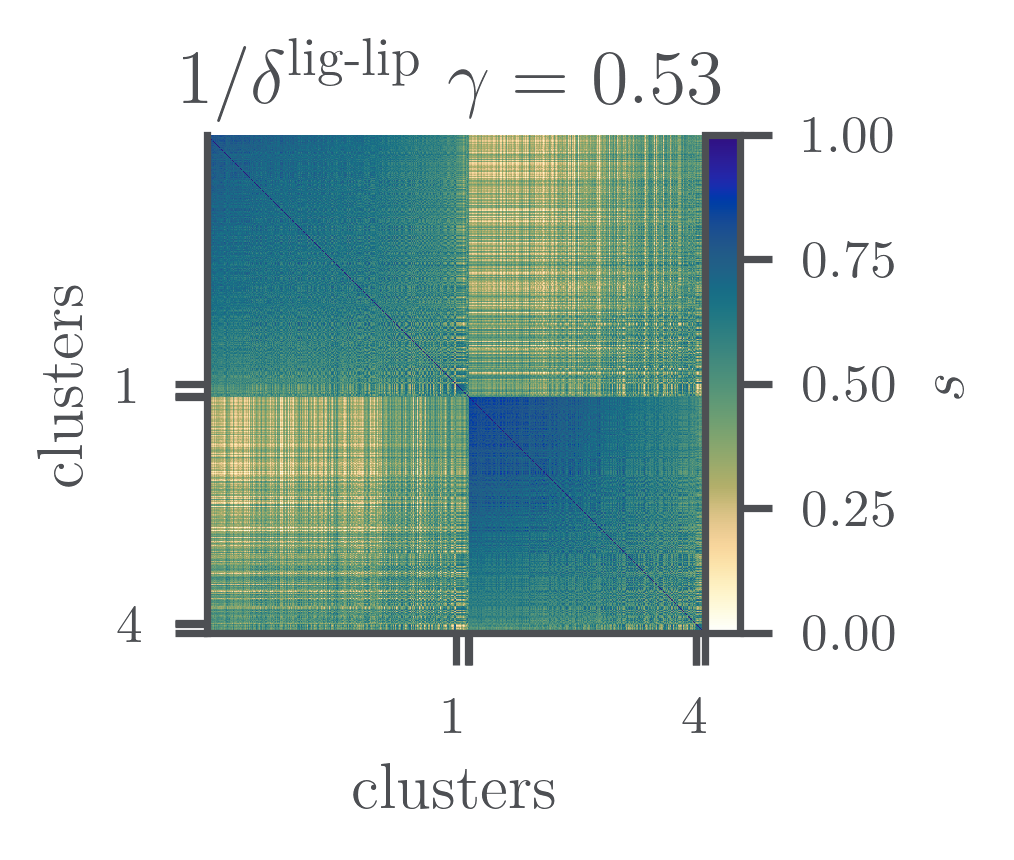}
        \caption{}
        \label{fig:A2A_liglipdistmatrix_invdist}
    \end{subfigure}
    \begin{subfigure}{0.4\linewidth}
        \includegraphics[width=\textwidth]{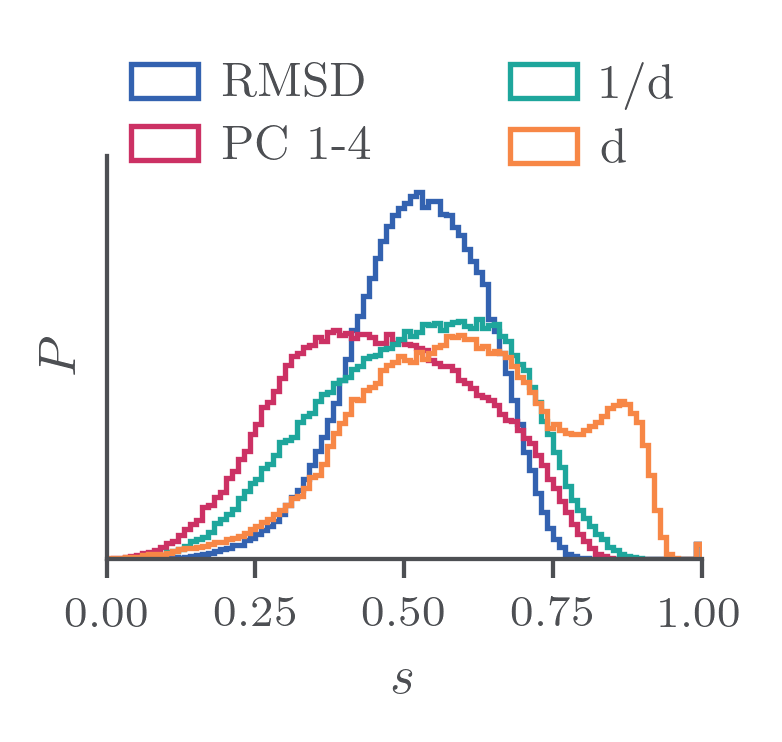}
        \caption{}
    \label{fig:A2A_ZMA_similarity_distribution}
    \end{subfigure}
    \caption{(a)-(b) Similarity block matrices from clustering lipid-ligand minimal distances and inverse lipid-ligand minimal distances with $\gamma=\text{median}(s_{ij})$. (c) Comparing similarity distributions of RMSD, conPCA, $d$ indicating ligand lipid distance $\delta_{\rm ligand-lipid}$ and $1/d$ indicating the inverse ligand lipid distance $1/\delta_{\rm ligand-lipid}$ based similarity measures.}
    \label{fig:A2A_liglip}
\end{figure}

\begin{figure}[H]
    \centering
    \begin{subfigure}{0.8\linewidth}
        \includegraphics[width=\textwidth]{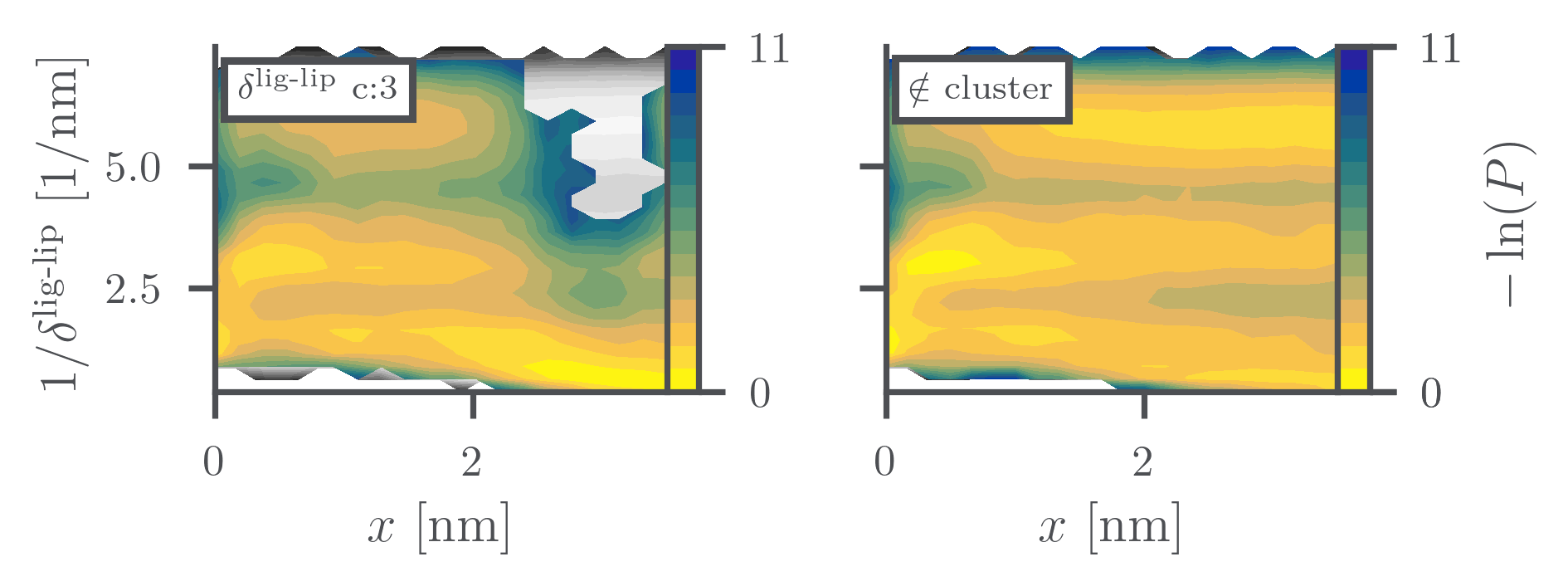}
    \end{subfigure}
    \begin{subfigure}{0.8\linewidth}
        \includegraphics[width=\textwidth]{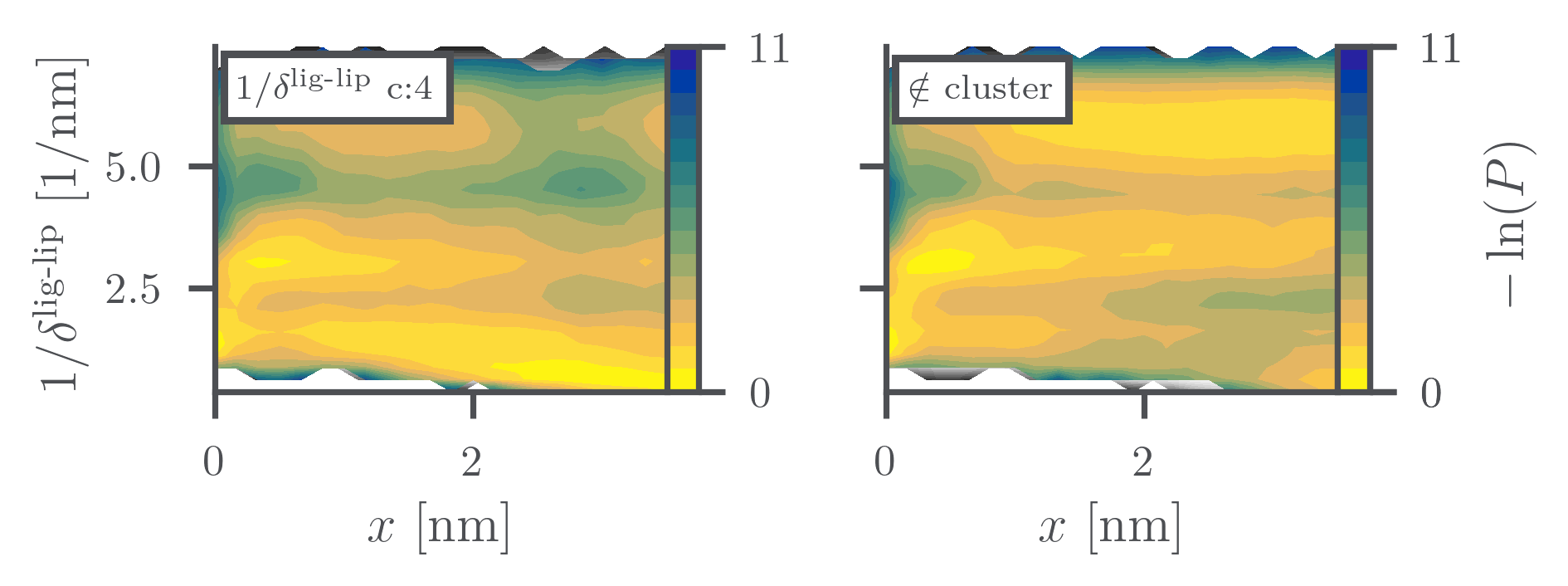}
    \end{subfigure}
    \begin{subfigure}{0.8\linewidth}
        \includegraphics[width=\textwidth]{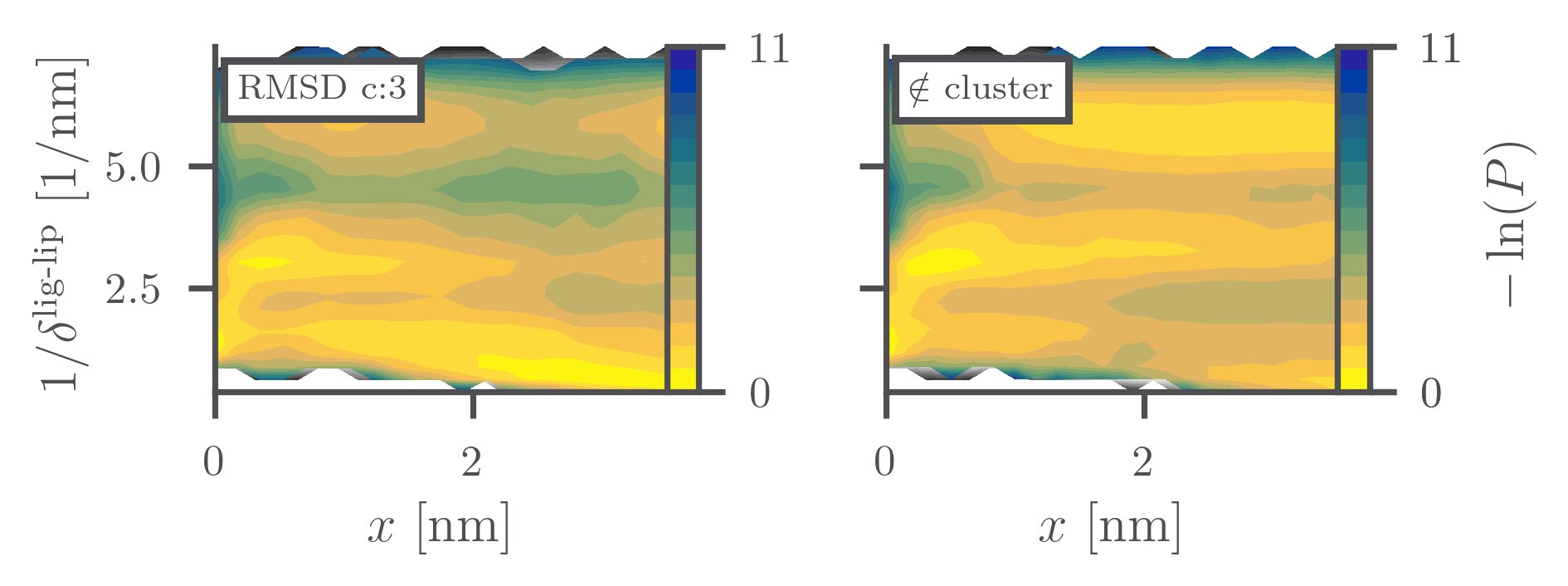}
    \end{subfigure}
    \begin{subfigure}{0.8\linewidth}
        \includegraphics[width=\textwidth]{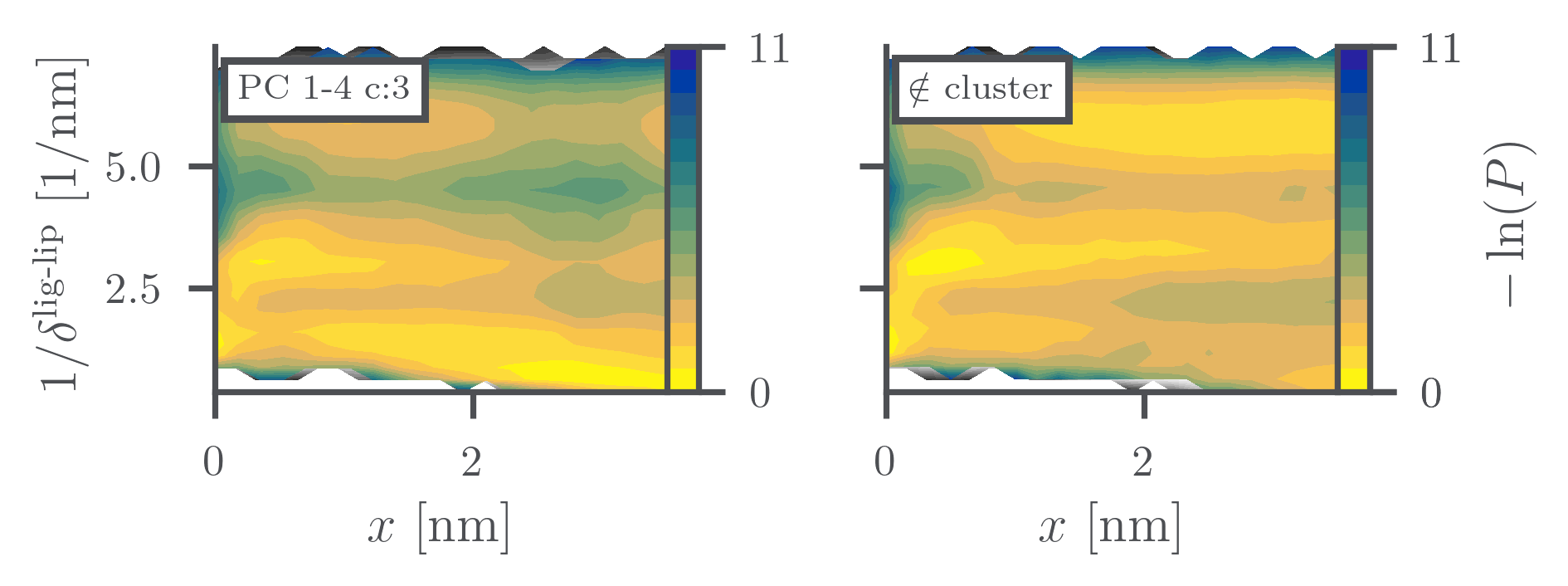}
    \end{subfigure}
	\caption{Inverse ligand lipid distance $1/\delta^\text{lig-lip}$ distribution over the pulling distance $x$ of all trajectories in grayscale, overlaid by different clustering results using $\gamma=\text{median}$ on the left and their complement on the right. The input feature and the cluster is indicated by the label.}
	\label{fig:SIA2Alipidligand_FEL}
\end{figure}

\subsubsection{Ligand-extracelular loop 2 (EL2) contacts}
\begin{figure}
    \centering
    \includegraphics[width=\linewidth]{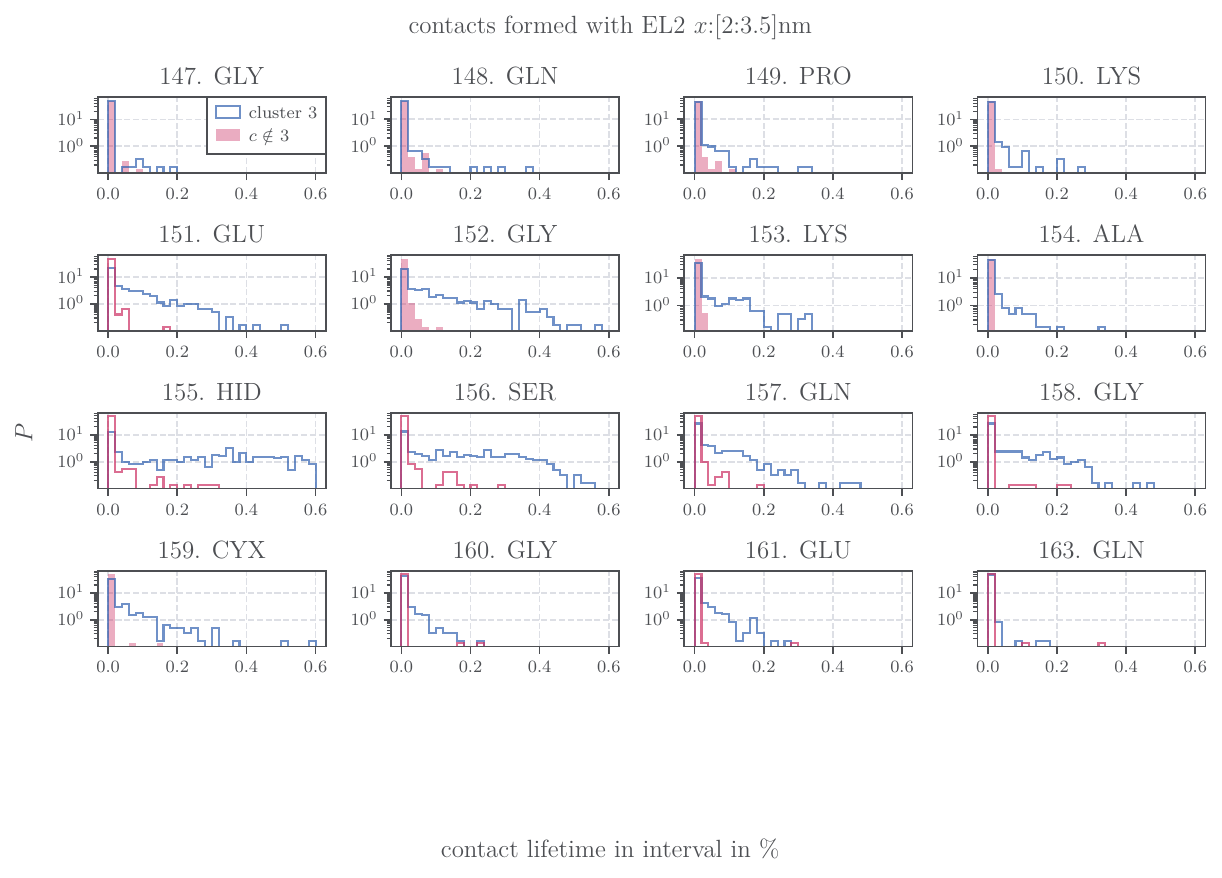}
    \caption{Comparing the population of EL2-ligand contact ($d<\SI{3.5}{nm}$) lifetimes (the number of frames the contact is formed in the pulling distance interval $x:2$-\SI{3.5}{nm} in \%) between cluster 4 from 1/$\delta^\text{lig-lip}$ $\gamma=0.53$ clustering (in blue) and its complement (in red). The solid histograms indicate a lifetime below $15\,\%$. 
    %The biggest difference in the distributions is visible for residues 153 (GLU), 154 (GLY), 157 (HID) and 158 (SER). 
    %\todo[inline]{SW to MJ: all residue numbers need to be subtracted by 2.}
    %Visual inspection of the residues during unbinding shows three behaviours:  1. the ligand sticks to the two residues during unbinding, 2. the ligand leaves the binding pocket and comes close to the residue during unbinding, but doesn't stick to them and 3. the ligand moves directly into the membrane. Combing this with the PCA eigenvector contributions where 1GLY and 160 GLY have the biggest contributions completes the image. Almost all ligands make a contact with 1 GLY during unbinding. While 160 GLY is sitting below 157 and 158.
    }
    \label{fig:contact_EL2_percent lipid ia}
\end{figure}
%\todo[inline]{What do you mean by "The solid histograms indicate a population $P<15\,\%$"? Does it mean low lifetime? How is lifetime defined?}
%\todo[inline]{Please add what these special residues relate to, so the reader knows how to interpret this.}
%\todo[inline]{In the title, please use ", $x\in [0,1]\,$nm" (no = and :).}

\newpage

\providecommand{\latin}[1]{#1}
\makeatletter
\providecommand{\doi}
  {\begingroup\let\do\@makeother\dospecials
  \catcode`\{=1 \catcode`\}=2 \doi@aux}
\providecommand{\doi@aux}[1]{\endgroup\texttt{#1}}
\makeatother
\providecommand*\mcitethebibliography{\thebibliography}
\csname @ifundefined\endcsname{endmcitethebibliography}
  {\let\endmcitethebibliography\endthebibliography}{}

\end{document}